# TEM analyses of unusual presolar silicon carbide: Insights into the range of circumstellar dust condensation conditions

S. A. Singerling[12], N. Liu[3], L. R. Nittler[4], C. M. O'D. Alexander[4], and R. M. Stroud[2]

Abstract

Presolar silicon carbide (SiC) grains in meteoritic samples can help constrain circumstellar condensation processes and conditions in C-rich stars and core-collapse supernovae. This study presents our findings on eight presolar SiC grains from AGB stars (four mainstream and one Y grain) and core-collapse supernovae (three X grains), chosen on the basis of μ-Raman spectral features that were indicative of their having unusual non-3C polytypes and/or high degrees of crystal disorder. Analytical transmission electron microscopy (TEM), which provides elemental compositional and structural information, shows evidence for complex histories for the grains. Our TEM results confirm the presence of non-3C,2H crystal domains. Minor element heterogeneities and/or subgrains were observed in all grains analyzed for their compositions. The C/O ratios inferred for the parent stars varied from 0.98 to ≥1.03. Our data show that SiC condensation can occur under a wide range of conditions, in which environmental factors other than temperature (e.g., pressure, gas composition, heterogeneous nucleation on pre-condensed phases) play a significant role. Based on previous μ-Raman studies, ~10% of SiC grains may have infrared (IR) spectral features that are influenced by crystal defects, porosity, and/or subgrains. Future sub-diffraction limited IR measurements of complex SiC grains might shed further light on the relative contributions of each of these features to the shape and position of the characteristic IR 11-μm SiC feature and thus improve the interpretation of IR spectra of AGB stars like those that produced the presolar SiC grains.

[1] Corresponding author sheryl.singerling.ctr@nrl.navy.mil, NRC Postdoc
[2] U.S. Naval Research Laboratory, Code 6366, Washington, DC 20375, USA
[3] Department of Physics, Washington University in St. Louis, St. Louis, MO 63130, USA
[4] Earth and Planets Laboratory, Carnegie Institution of Washington, Washington, DC 20015, USA

# 1. INTRODUCTION

Presolar circumstellar grains found in primitive meteorites and interplanetary dust particles are the oldest solids available for study, predating solar system formation by up to a few billion years (Heck et al. 2020). These micron- to nanometer-sized dust particles vary in size and mineralogy (e.g., diamond, silicon carbide, graphite, oxides, silicates), but share the common characteristic of having very anomalous isotopic compositions. The grains' O, C, N, Si, or other isotopic compositions can differ by orders of magnitude from those of materials that formed in the solar system. These compositions reflect the formation of the grains by condensation from the products of stellar nucleosynthesis in environments ranging from the envelopes of asymptotic giant branch (AGB) and red giant branch stars to the ejecta of novae and core-collapse supernovae (CCSNe). Presolar grains represent the only physical samples of these circumstellar environments available to us, and, as such, they provide constraints on condensation of dust in circumstellar environments, as well as stellar nucleosynthesis and evolution. The grain morphologies, crystal structures, minor element and subgrain contents, which can be determined from transmission electron microscopy (TEM) studies of the grains, reflect grains' individual circumstellar condensation conditions. Such detailed information can guide models for interpretation of astronomical observations of the infrared (IR) spectra of dust grains and for better constraining dust condensation processes in different circumstellar environments.

Silicon carbide (SiC) grains are the most well-studied type of presolar grain, and they are divided into several groups — mainstream (MS), Y, Z, X, C, U, PNG, AB — largely based on differences in C, N, and Si isotopic compositions that reflect different classes of progenitor stars (Nittler & Ciesla 2016 and references therein). Stellar environments for the presolar SiC groups include: C-rich AGB stars with close-to-solar or higher metallicities (MS grains); C-rich AGB stars with sub-solar metallicity and/or greater than solar mass (Y and Z grains); CCSNe for X, C, and possibly the majority of U grains (Xu et al. 2015); and novae and CCSNe for PNG grains (e.g., Nittler & Hoppe 2005; Liu et al. 2016). The origin of AB grains is more ambiguous, and their proposed stellar sources include CCSNe, J-type C-stars, and novae (Liu et al. 2017a; Schmidt et al. 2018; Hoppe et al. 2019). MS grains are the most common isotopic group, making up about 90% of all SiC grains, followed by AB (~5%), Y (~1-5%), Z (~1-5%), X (~1-2%), C (<1%), U (<1%), and PNG (<1%) grains.

The crystalline structures of SiC display polytypism, a one-dimensional variant of polymorphism with differences in the stacking order of Si-C pairs. The polytypes form superlattices with cubic (C), hexagonal (H), or rhombohedral (R) crystal structures; the cubic polytype is also referred to as β-SiC, whereas all non-cubic polytypes are collectively referred to as α-SiC. Over 250 variations have been identified in synthetic SiC that were grown at differing temperatures, pressures, etc. (Fisher & Barnes 1990; Bechstedt et al. 1997; Shiryaev et al. 2011). Despite this potential diversity, presolar SiC shows remarkable consistency in polytype. Using TEM, Daulton et al. (2003) analyzed 508, 0.32–0.70 μm-sized presolar SiC grains isolated from the Murchison meteorite for their polytypes, and determined that presolar SiC grains are predominantly cubic 3C (~80%), hexagonal 2H (~3%), or intergrowths of the 3C and 2H polytypes (~17%). Similarly, most of the IR astronomical spectra of circumstellar SiC, associated with the "11-μm" feature, are consistent with a distribution of sub-micron-sized 3C (i.e., β-SiC) grains (Speck et al. 1999; Clement et al. 2003).

Coordinated isotopic and structural data have the potential to identify differences in circumstellar condensation conditions among different stellar environments. However, the isotopic compositions of the SiC grains were not measured in the Daulton et al. (2003) study. Consequently, the grains could not be assigned to a group, but are presumed to belong primarily to the MS group, given that it is the most common. Subsequent studies have determined the polytypes of 36 non-MS grains, including 22 X, 10 AB, two Z, one Y, and one C (Stroud et al. 2004; Daulton et al. 2006, 2009; Hynes 2010; Hynes et al. 2010; Liu et al. 2017b; Gyngard et al. 2018; Kodolanyi et al. 2018). These investigations found 3C, 2H, 6H, and 15R polytypes amongst the grains, with 15 grains that had multiple domains and were actually intergrowths of more than one polytype (i.e., 3C-2H, 3C-2H-15R, 3C-6H). From the limited data set of available non-MS grain crystal structures, the 3C and 2H polytypes are the most common, which implies generally similar condensation conditions across both the parent AGB stars and CCSNe for the different SiC groups, although there are exceptions.

In addition to crystal structure, minor element (Al, Mg, and N) and subgrain compositions can also provide useful constraints on formation conditions. Thermodynamic equilibrium (i.e., condensation) calculations predict the formation of phases such as carbides (e.g., TiC, $Fe_3C$, ZrC, MoC), nitrides (e.g., AlN, TiN), and sulfides (e.g., MgS, CaS) at certain temperatures under specific pressures and C/O ratios in circumstellar environments, including both AGB stars and CCSNe (Lodders & Fegley 1995; Sharp & Wasserburg 1995; Hoppe et al. 2001). Such modeling is pertinent for subgrains that formed before or concurrent with the SiC and were incorporated into their host grains. Lodders & Fegley (1995) and Hoppe et al. (2001) also provide information on the formation of phases that can form solid solutions with SiC, which is relevant for subgrains or minor element heterogeneities that formed as a result of exsolution from SiC on cooling.

There have been only a limited number of TEM studies of minor element and subgrain compositions in presolar SiC grains. These studies have mostly focused on grains from CCSNe (X and C grains) (Stroud et al. 2004; Hynes et al. 2010; Liu et al. 2017b; Gyngard et al. 2018; Kodolanyi et al. 2018), although some MS grains (Hynes et al. 2010; Liu et al. 2017b) and AB grains (Hynes 2010) have also been studied. More detailed investigations of subgrains in C-rich presolar grains have been performed on graphite grains, in which subgrains are common; Bernatowicz et al. (1996) observed metal carbides in one-third of the 67 graphite grains that they studied, from a variety of stellar sources, and Croat et al. (2003) noted a high abundance (25–2400 ppm) of TiC subgrains in the CCSN grains that they analyzed. The latter study also observed Fe,Ni-bearing (kamacite/taenite) metal subgrains.

Given the rarity (few %) of the non-MS presolar SiC grains and the time-consuming process for identifying them, Liu et al. (2017b) developed a new screening tool for locating presolar SiC grains from rare isotope groups and/or with unusual crystal structures using coordinated scanning electron microscopy (SEM) energy dispersive X-ray (EDX) spectroscopy and micro-Raman spectroscopy (hereafter, EDX-μ-Raman). In Raman spectroscopy of SiC, the transverse optical (TO) modes, which measure the stretching of the Si-C bond pair, vary in number and position depending on SiC polytype. The peak widths of the modes are affected by the presence of crystal defects, such as stacking faults that are common to the 3C polytype, and compositional variations. Given the dependency of peak positions on polytype, the EDX-μ-Raman screening tool allows for rapid, non-destructive identification of non-3C SiC polytype candidates, as Liu et al. (2017b) confirmed for two of the three presolar SiC grains that they analyzed with TEM. Based on μ-Raman spectra, Liu et al. (2017b) found that ~10% or more of presolar SiC grains exhibit non-3C bulk structures (i.e., grains that were not entirely composed of 3C crystal domains), after correcting for the fact that X grains were overrepresented in the study. The 11-μm IR feature associated with SiC dust in astronomical spectra also derives from the atomic vibrations along Si-C bonds, and includes both the TO and longitudinal optical (LO) modes.

Thus, grains that exhibit distinct Raman spectra could also give rise to IR spectra that deviate from the predominant 3C-SiC spectral distribution. The goals of this study are to: (1) explore the structural and elemental characteristics that may influence the Raman spectra, (2) gain insights into dust formation under a wide range of conditions in the different circumstellar environments, and (3) provide guidance for better constraining the interpretation of astronomical IR spectra.

## 2. METHODS

### 2.1 Focused ion beam sample preparation

Eight presolar SiC grains were selected for study that had previously been pressed into high purity Au, then analyzed for structural information by μ-Raman spectroscopy, and finally analyzed for their C, N and Si isotopic compositions by NanoSIMS (Liu et al. 2017b). Electron transparent sections of the grains were prepared by the in situ lift-out technique with an FEI Helios DualBeam FIB-SEM at the U.S. Naval Research Laboratory (NRL). The eight grains were MS grains M2-A1-G312, M2-A1-G619, M2-A1-G620, and M2-A1-G648; Y grain M2-A1-G670; and X grains M1-A4-G506, M2-A1-G674, and M2-A2-G1036. Hereafter, the M#-A# portion of the grain name is excluded. The FIB preparation steps were: (1) depositing a protective C strap over the grain with an FEI Multichem Gas Injection System, (2) cutting a lamella of the grain from the supporting Au foil with a $Ga^+$ ion beam, (3) extracting the lamella with an EasyLift NanoManipulator, (4) mounting the lamella onto a TEM Cu half grid using Pt and/or C deposition, and (5) thinning the lamella to electron transparency (i.e., 80–100 nm) with progressively smaller ion beam currents and accelerating voltages. The working distance was 4 mm. Secondary electron imaging with the electron beam was performed at 5 kV and 1.6 nA. Conditions for the ion beam milling varied — 30 kV and 25 pA or 40 pA for imaging, 30 kV and 2.5 nA for milling, 30 kV and 40 pA or 80 pA for deposition, 30 kV and 0.23 nA for initial thinning, and 15 kV and 80 pA down to 20 pA for final thinning. Note that given the high mobility of Au under the ion beam, several grains contain Au contamination as a consequence of FIB-sample preparation. The grains with Au contamination are labeled as such in HAADF images.

### 2.2 Transmission electron microscopy

We analyzed the presolar SiC grains for structural and compositional information on two TEMs at NRL — a JEOL 2200FS TEM and a Nion UltraSTEM-200X, both at 200 kV with double tilt holders. On the JEOL, a Gatan OneView camera was used to collect bright field (BF) and "dirty" dark field (DDF) images, as well as selected area electron diffraction (SAED) patterns. To collect the DDF images, we placed a high contrast objective aperture over a diffraction spot in diffraction mode, then switched to imaging mode and adjusted the brightness to get a representative DDF image. For imaging on the JEOL, the condenser lens aperture 2 (100 μm) was used with spot size 1. The selected area aperture size varied depending on the size of the grain, and most SAED patterns were collected at a 50 cm camera length. On the Nion, a high angle annular detector with an inner angle of 66 mrad was used to collect scanning TEM (STEM) high angle annular dark field (HAADF) images. Gatan Digital Micrograph was used to process images collected on both the JEOL and the Nion microscopes. We preferentially used Calidris CRISP software in conjunction with the phase identification function PhIDO to index our SAED patterns and determine their polytypes. For patterns that CRISP was unable to index with high certainties, we collected *d*-spacing values and angles using Gatan Digital Micrograph DiffTools and compared the measured values with theoretical ones. In both cases, the SAED patterns were checked against the following polytypes: 3C, 2H, 4H, 6H, 8H, 10H, 14H, 15R, and 19H. Data on the crystal structures of the polytypes are summarized in the appendix (Table A3). For each crystal domain in the presolar SiC grains with non-3C or non-2H polytypes, we collected a minimum of three SAED patterns with known goniometer tilts (three for G619, six to seven for each 8H domain in G506, five for the 10H domain in G674, and three for the 14H domain in G674). Grain sizes were calculated from the TEM BF images of the extracted cross-sections using the geometrical mean method, in which we measured the diameter by bisecting the projected area of the grain in multiple locations and then taking the *n*th root of the product of those diameters (see Table A1 for data).

On the Nion, a Bruker Xflash SDD system was used to collect EDX maps and spectra at an average current of 60 pA. EDX maps were collected for 15–20 mins at a resolution of 4 nm per pixel, and individual spectra of subgrains were collected for 1.5–2 mins. For EDX data processing, we used the Bruker ESPRIT 2.0 software to calibrate spectra, correct the background, and check the deconvolution. Quantification of elemental abundances was carried out using the Cliff-Lorimer method with detector-specific computed *k* factors. The following elements were included in the quantification only for deconvolution purposes, as they represent contamination from sample preparation, SIMS measurements, or components within the TEM or sample holder: O, Cu, Ga, and Cs. Zirconium was included only for deconvolution purposes in spectra that did not show significant $L_{\alpha1}$ and $L_{\beta1}$ energy peaks for Zr. For subgrain spectra which showed Zr L peaks, we corrected for Zr contamination from the EDX detector by collecting a control spectrum from adjacent SiC and subtracting the Zr in the control spectrum from the Zr in the respective subgrains.

## 3. RESULTS

The C, N, and Si isotopic compositions were used to assign each grain to an isotopic group. Figure 1 depicts the C and N (Fig. 1a) and Si (Fig. 1b) isotopic compositions of the eight grains studied, along with literature envelopes (data from Stephan et al. 2020) for the groups to which the grains belong (i.e., MS, Y, and X). In the following sections, we describe the textural and compositional characteristics of the eight grains studied. We group them by their isotopic group (MS, Y, and X) and further subdivide the MS grains by their polytypes — those that contain only the 3C polytype (i.e., 3C bulk structures) and those that contain polytypes other than 3C (non-3C bulk structures).

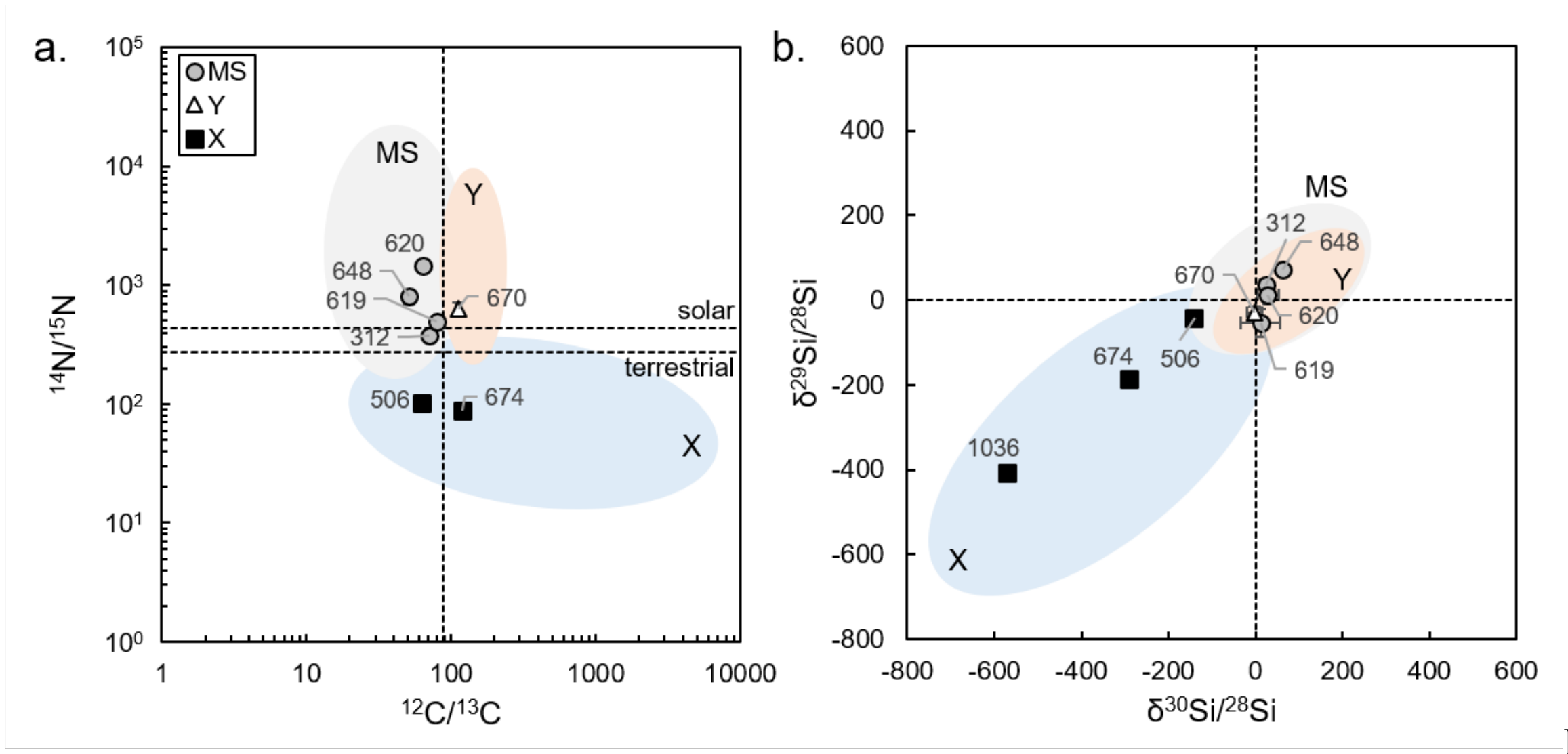

**Figure 1.** C vs. N (a) and Si (b) isotopic compositions of the eight presolar SiC grains studied. G1036 is not included in (a) as its N-isotopic composition was not available. The envelopes represent literature values for the MS (gray), Y (orange), and X (blue) groups, sourced from Stephan et al. (2020). The dashed lines indicate the terrestrial (a and b) and solar (a only, from Marty et al. 2011) values. Symbol colors and shapes indicate the groups (gray circles = MS, white triangle = Y, black squares = X). Error bars are 1σ, most of which are smaller than their symbols.

Table 1 summarizes the textural information for the grains along with their isotopic and Raman data. Table 2 lists the minor element compositions of the SiC. Table 3 summarizes the subgrain textural and compositional data, both as absolute compositions and elemental atomic ratios. In Table 3, the subgrain compositional data are represented as elemental atomic ratios only. This is because the EDX analyses likely represent subgrain as well as adjacent SiC compositional information, rather than just the subgrain composition. Additionally, the Cliff-Lorimer quantification is strongly dependent on thickness and density, and given the unknown contribution of SiC and subgrain to the overall analyses, these two parameters are not well constrained. Figures 2–6 display TEM bright field images, STEM HAADF images, EDX maps, and/or SAED patterns of the grains. Additional data are presented in the appendix (Figs. A1–A10 and Tables A1–A7).

### 3.1 Mainstream grains

#### *3.1.1 3C bulk structures*

The MS grains with Raman spectra consistent with 3C bulk structures (i.e., only crystal domains of the 3C polytype) include G312, G648, and G620 (Fig. 2). We included these grains in this study to act as “normal controls” for comparison with any non-3C bulk structure MS grains. For each grain in this study, SAED patterns and DDF imaging were used to determine the number of crystal domains present. Representative DDF images of the grains are included in the appendix (Fig. A1). G648 consists of only one crystal domain, which indexes to the 3C polytype. G620 is composed of two crystal domains; one of these indexes to 3C, but we were unable to index the other domain given a poor TEM calibration from that session. We have tentatively assigned the grain’s polytype as 3C, given its Raman spectrum (see section 4.1). G312 appears to be composed of three crystals, all of which index to 3C with varying levels of stacking disorder. However, similarities in the orientations of the grains (Fig. A1), the presence of high-density stacking faults in the lower portions of the middle and rightmost fragments (linear features in Fig. 2a), and the appearance of the grain on the Au foil (Fig. A2) all imply that the three crystals are from a single grain that was fractured during sample preparation when pressed into the supporting Au foil.

The grains range in size (calculated from the geometrical mean method as described in section 2.2) from 710 nm to 1160 nm and in shape from circular to highly elliptical. Crystal defects, specifically stacking faults, are not present in G648 or G620, but, as previously mentioned, are present in G312. Voids, best observed in the STEM HAADF images of the grains (Fig. 2b, 2e), are present in low abundance in G312 and G620. The voids are 10s of nm in size (see Table A1 for detailed size data) and are located along grain boundaries (G620) or have a random distribution unassociated with structural features (G312). Unfortunately, G648 was lost during sample exchange, so only limited data (BF images and SAED patterns) are available for this grain.

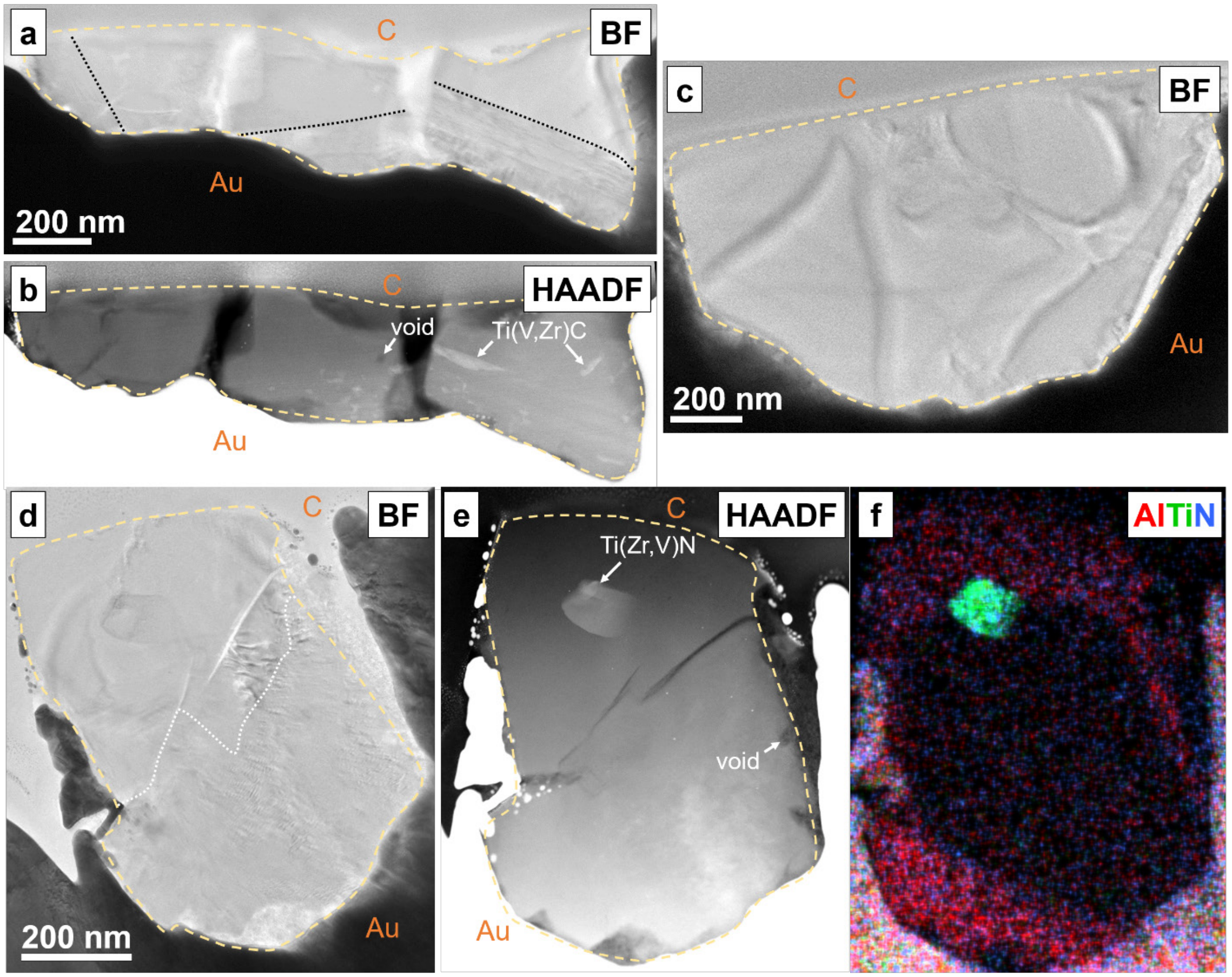

**Figure 2.** BF images, STEM HAADF images, and an EDX map of MS grains with only 3C domains: (a–b) G312, (c) G648, and (d–f) G620. The dashed lines outline the SiC grains, and the stippled white line in (d) shows the boundary between crystal domains. The Au foil, supporting the SiC, and the protective C strap are labeled. Small, bright white material within the grains' HAADF images (b, e) is Au contamination from FIB sample preparation Large TiC and TiN subgrains are present in G312 (b) and G620 (e), respectively. Regions of high-density stacking faults in G312 (a) are indicated with the stippled black lines in the lower portions of the middle and rightmost fragments and the left portion of leftmost fragment. The HAADF image of G312 (b) illustrates that these highly defected regions contain a higher

abundance of the TiC subgrains. The EDX map of G620 (f) shows the presence of Al,N-rich SiC (red-blue) near the edges of the grain as well as a single, large Ti(Zr,V)N subgrain.

Impurity atoms are incorporated into presolar SiC as distributed substitutional and interstitial defects within the SiC lattice, as well as in subgrains (also called “inclusions” in the literature). The minor elements Al and N are commonly distributed throughout the lattice, but occasionally occur as distinct subgrains. Table 2 and Figure A3 allow for a comparison of the minor element contents of each grain. Table 2 includes minor element compositional data for representative regions of SiC, whereas Figure A3 presents spectra of the overall grains. Other elements (e.g., Ti, Zr, V, Fe, Ni) are mostly present in the form of subgrains. Given the small sizes of the subgrains, many do not extend through the entire thickness of the FIB section (~80–100 nm). Contributions from SiC are expected in the spectra of the subgrains, but by comparing the relative abundances of elements in the analyses of the subgrain with adjacent SiC, we can get a better sense of any enrichments or depletions of C, Si, and the minor elements (Al, N).

G312 shows a significant difference in Al and N contents between the stacking fault-rich and stacking fault-poor portions of the grain (Table 2; Fig. A4c). The areas of high-density stacking faults show greater N (0.76 ± 0.06 at.%) and Al (0.26 ± 0.05 at.%) contents compared to the less disordered portions of the grain, which contain 0.28 ± 0.02 at.% N and 0.15 ± 0.04 at.% Al. Titanium-bearing subgrains with minor amounts of V and Zr, and Al,N-bearing subgrains (not visible in Fig. 2b) are also present in G312. The Ti- and Al,N-bearing subgrains range in size from <5 nm to 60 nm, and in shape from equant to elongate. Although the G312 subgrains are mostly unassociated with voids, they are predominantly located in the high-density stacking fault regions of the grains (Fig. 2a) and show a preferred orientation of their elongation direction parallel to the defects.

Comparing EDX data from the Ti-bearing subgrains with the adjacent SiC in G312 reveals N enrichments (as much as +1.5 at.%; enrichment and depletion values are relative to adjacent SiC) and both C enrichments (as much as +13.8 at.%) and depletions (as much as -3.2 at.%). Small depletions in C do not exclude Ti carbide (TiC) as a candidate for the phase of the Ti-bearing subgrains, especially since transition metal carbides regularly show stoichiometric variations (e.g., Williams 1988). The crystal structures of transition metal carbides often contain vacancies in the C sites, so C depletions, as compared to SiC, are not unexpected. Given the small N enrichments, the Ti-bearing subgrains in G312 are likely TiC, although TiCN is also a possibility. The TiC subgrains in G312 do not have large heterogeneities in their Zr/Ti and V/Ti ratios (Table 3; Fig. A4d). Two Al,N-bearing subgrains were identified in EDX maps of G312 (Fig. A4b). Consistent with their being AlN, these subgrains have Al/N ≈1. This is in contrast with the Al,N-rich regions in other MS grains (G620 and G619, discussed below), which are larger and have Al/N ≤0.7.

G620 contains Al,N-rich SiC regions (4.6 ± 0.2 at.% N and 2.3 ± 0.2 at.% Al) in the outer portions of the grain that outline a euhedral, Al,N-poor SiC core (0.55 ± 0.07 at.% N and Al below detection limit) (Fig. 2f). G620 also contains one large (100 nm) Ti-bearing subgrain, which also contains minor amounts of Zr and V. The subgrain is equant but wedge-shaped as illustrated by the diffuse appearance of its lower-right boundary in Figure 2e. The subgrain is not associated with voids, although it is located at the interface between the Al,N-rich and Al,N-poor SiC regions of the grain. Comparing EDX data from the subgrain with the adjacent SiC reveals a large N enrichment (+5.5 at.%) and C depletion (-10.7 at.%), implying that the subgrain is likely a Ti nitride (TiN) rather than TiC, although we cannot exclude the possibility of TiCN.

### *3.1.2 Non-3C bulk structure*

Consistent with predictions from the μ-Raman spectra, MS grain G619 contains a non-3C bulk structure (i.e., at least one crystal domain that is not the 3C polytype) (Fig. 3). G619 is predominantly one crystal domain, which indexes to the 4H polytype. We cannot use the DDF images of G619 for deducing crystal domains, as the varying brightness in the DDF images is dominated by curtaining effects (i.e., non-uniform thicknesses across the section resulting from different FIB-sputtering rates at different locations) rather than differences in the crystallographic orientations of domains (Fig. A1). However, SAED patterns require intergrowths of the 2H and 4H polytypes in one region of the grain, hence its polytype designation of 2H-4H.

Figure 3i–iii illustrates three SAED patterns collected from G619. The dashed, orange circles in Figure 3a indicate from where in the grain the SAED patterns were collected. Patterns (i) and (ii) come from the same location but at different goniometer tilts and index to 4H, whereas pattern (iii) was collected in a different location in the grain. Pattern (iii) includes spots with intensities (dashed, orange circles) inconsistent with 4H SiC alone. To match the intensities of those spots, some kind of intergrowth is required. Indexing of those spots (Tables A4 and A5) and simulated electron diffraction patterns are consistent with intergrowths of either 4H SiC and 2H SiC, 4H

SiC and AlN, or 4H SiC and TiC. However, a 2H-4H intergrowth is the most appropriate polytype designation for G619, consistent with the Raman data. AlN is less likely, given that the Al,N-rich regions are actually Al,N-rich SiC rather than pure AlN, as will be discussed below. TiC is also less likely given that the individual TiC subgrains are too small to contribute such well-defined diffraction spots, and the simulated diffraction pattern for TiC is not a perfect match to the experimental one.

G619 is 670 nm in size and roughly circular in shape. Crystal defects, specifically stacking faults, are not observed in the grain. The vertical linear features in G619 (Fig. 3a–b) are likely curtaining produced during FIB-section preparation and are not inherent to the grain itself, as pointed out earlier. However, other strain contrast features are present in the grain, most visibly in the lower left of the BF image (Fig. 3a). Numerous voids (>50) are present and range in size from <10 nm to 10s of nm (see Table A1 for detailed size data).

G619 contains a complex texture of Al,N-rich (3.0 ± 0.2 at.% N and 2.1 ± 0.2 at.% Al) and Al,N-poor SiC (0.52 ± 0.08 at.% N and 0.11 ± 0.05 at.% Al) (Fig. 3c, 3f). The innermost Al,N-rich bands outline a euhedral, Al,N-poor SiC core. Numerous SAED patterns collected across G619 did not reveal a significant difference between the Al,N-rich and Al,N-poor SiC; however, the selected area aperture used was larger than the Al,N-rich regions. It may be that Al,N-rich and Al,N-poor SiC are composed of different polytypes (i.e., 4H for Al,N-poor SiC and 2H for Al,N-rich SiC), but the relative size of the crystal domains and the aperture limited the ability to observe this. As mentioned previously, we were unable to use DDF imaging to determine the location and number of crystal domains. Interestingly, there seem to be multiple generations of bands separated by Al,N-poor SiC with an apparent association between the Al,N-rich SiC bands, voids, and subgrains (Fig. 3b–c, 3e–f).

The subgrains are mostly Ti-bearing, with minor amounts of V, Fe, Zr, Mo, and/or Ni, and range in size from <5 nm to 30 nm. Many of the Ti-bearing subgrains are elongate with their direction of elongation being perpendicular to the AlN bands (Fig. 3e–f); however, some are equant (Fig. 3d). Comparing EDX data from the Ti-bearing subgrains with the adjacent SiC reveals N enrichments and both C enrichments and depletions; however, the subgrains are surrounded by the Al,N-rich SiC, making it difficult to discern if the N enrichments are in fact from the Al,N-rich SiC, given that the control SiC spectra were taken in Al,N-poor regions. Since the high magnification EDX map in Figure 3f does not show concentrated N in the subgrains, we argue that the N enrichments are, in fact, largely from the Al,N-rich SiC itself, although this does not discount the presence of elevated N abundances in the subgrains. As with G312, the slight C depletions are consistent with
transition metal carbides. Hence, the Ti-bearing subgrains are best identified as TiC, again with another possibility being TiCN. In addition to the TiC subgrains, G619 also contains an anhedral, equant, Fe-bearing subgrain (subgrain 3 in Table 3; Fig. A5), located on the edge of a void, and an equant CaS subgrain adjacent to a void and a TiC subgrain (Fig. 3d; EDX map showing CaS in Fig. A5). The Fe-bearing subgrain is likely Fe metal given its large Si (-4.5 at.%) and C depletions (-9.7 at.%), which preclude the possibility of Fe carbide or silicide. We cannot rule out Fe oxide as a possible candidate, given that O cannot be reliably determined by EDX owing to contamination of SiC during FIB sample preparation. However, for the C-rich parent stars of MS grains, Fe oxide is not an expected condensate based on theoretical calculations.

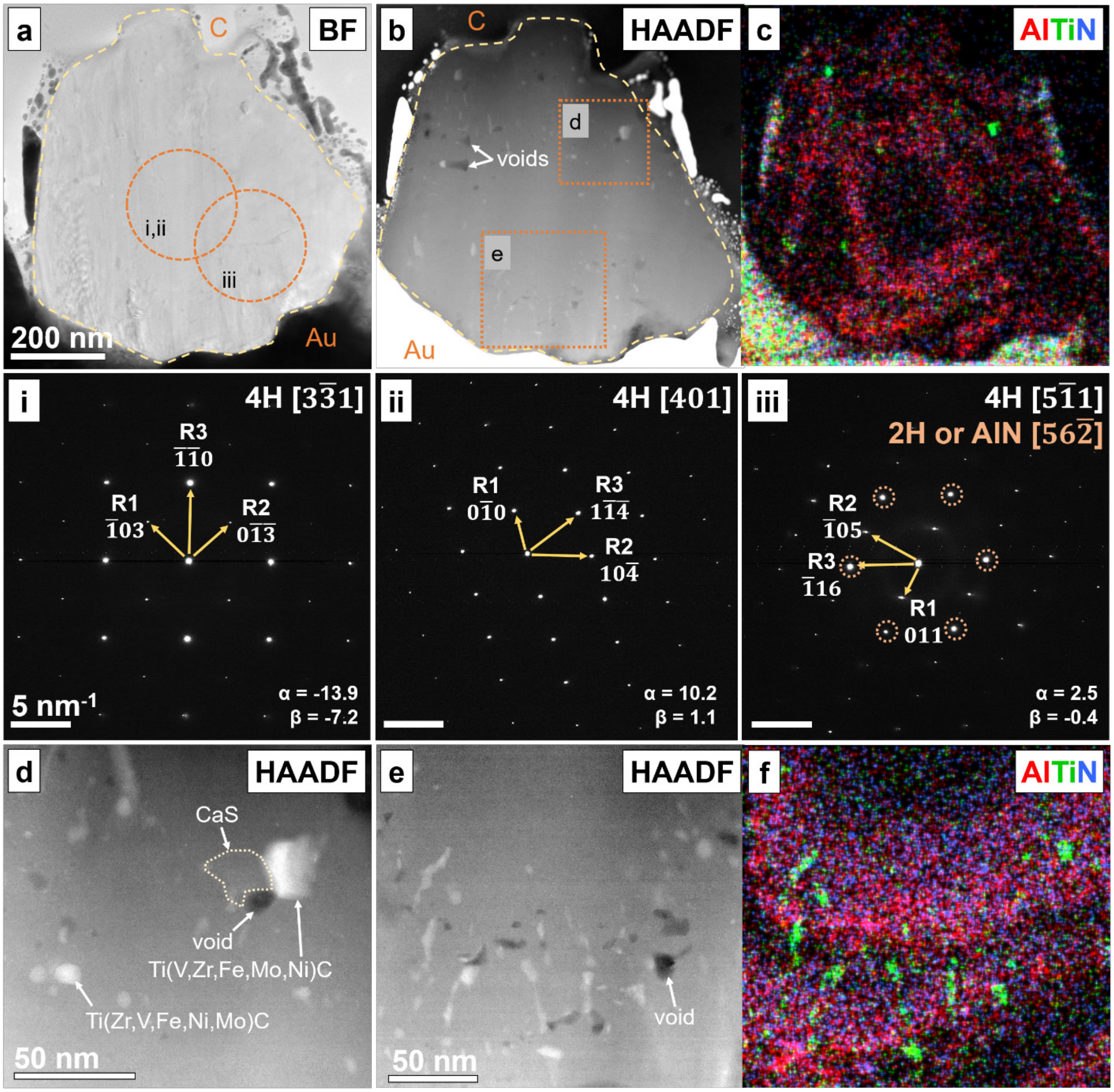

**Figure 3.** BF image (a), STEM HAADF images (b, d–e), EDX maps (c, f), and SAED patterns (i–iii) of the MS grain with a non-3C bulk structure, G619. The dashed lines outline the SiC grain. The Au foil, supporting the SiC, and the protective C strap are labeled. The dashed orange circles in the BF image show the locations where SAED patterns (i–iii) were collected. Patterns (i) and (ii) were collected at different goniometer tilts. Pattern (iii) indicates an intergrowth of 4H SiC and 2H SiC, AlN, and/or TiC. The higher intensity diffraction spots, indicated by the dashed orange circles, represent locations where 4H SiC and the other structures overlap. Numerous subgrains are present (d, e) and are predominantly TiC, although one CaS subgrain was observed and is outlined for clarity in (d). The EDX maps show the presence of Al,N-rich SiC bands (red-blue) alternating with Al,N-poor SiC (c, f). Higher abundances of TiC subgrains and voids are associated with the Al,N-rich SiC (e–f).

### 3.2 Y grain

G670 is the only Y grain studied (Fig. 5) and is composed of two crystal domains (Fig. A1), which both index to the 3C polytype. The SAED patterns included in Figure 4 illustrate the orientation relationship between the two domains, with the $(\bar{2}02)$ plane in the one domain parallel to the $(022)$ plane in the other domain. The grain is 640 nm in size and slightly elliptical in shape. Crystal defects, specifically several stacking faults, are present, but do not occur in the high densities observed in other grains (Fig. 4a, bottom tip of the grain). Several voids (>5) are also present (Fig. 4b–c) and range in size from 10s of nm to ~100 nm in size (see Table A1 for detailed size data). The Al and N abundances in G670 are very low with no detectable spatial heterogeneity (Table 2; Fig. A3). G670 contains four elongate, Ti-bearing subgrains, varying in size from 20 nm to 50 nm. Only one subgrain contains other transition metals, specifically minor amounts of V. Three of the four subgrains are associated with voids. Comparing EDX data from the subgrains with adjacent SiC reveals no N enrichment and both C depletion (as much as -2.4 at.%) and enrichment (as much as +5.3 at.%), implying that the subgrains are likely TiC.

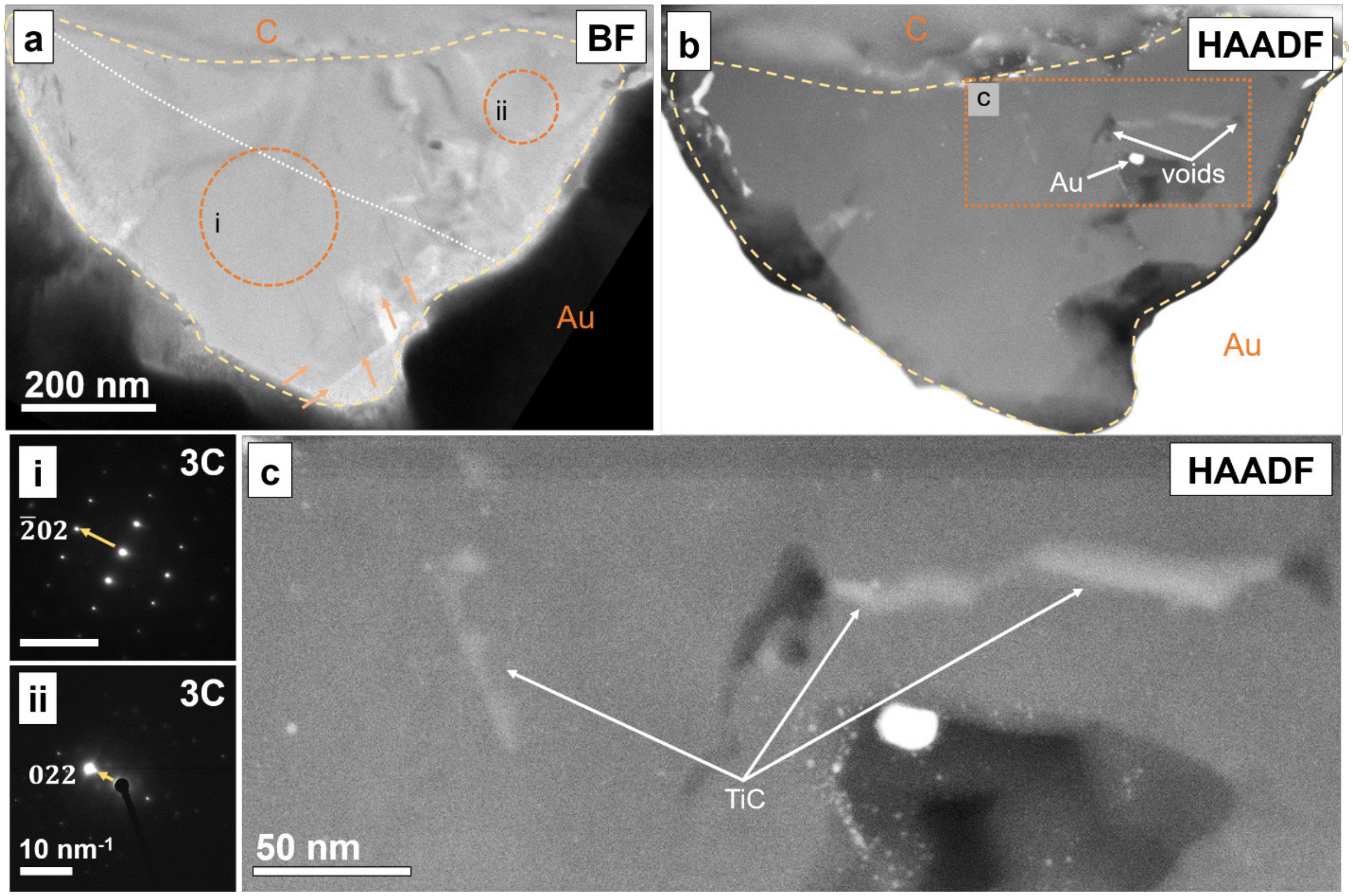

**Figure 4.** BF image (a), SAED patterns (i–ii), and STEM HAADF images (b, c) of Y grain G670. The dashed lines outline the SiC grain. The Au foil, supporting the SiC, and the protective C strap are labeled. Bright white material within the grain's HAADF images (b–c) is Au contamination from FIB sample preparation. G670 contains two crystal domains, with their contact indicated by the stippled white line in (a). The SAED patterns of the two domains are also included (i, ii), and the locations where they were collected are shown in (a) by the dashed orange circles. The two domains are both 3C and have the following orientation relationship: $\bar{2}02//022$. Orange arrows in the lower portion of (a) indicate and are parallel to stacking faults. Several TiC subgrains are present which are elongate and often associated with voids (b, c).

### 3.3 X grains

The three X grains studied are G506, G674, and G1036 (Fig. 5). All three grains are complex intergrowths with multiple crystal domains (greater than seven for G506, >13 for G674, and four for G1036; Fig. A1). The locations of the domains are indicated by stippled white lines within the images of the grains (Fig. 5a, 5d, 5h). For grains containing multiple polytypes, we represent these as polytype-polytype, with each polytype listed occurring at least once in the grain in question. The polytypes in G506 include 3C and 8H (3C-8H), and those in G674 include 3C, 2H, 10H, and 14H (3C-2H-10H-14H), although we did not identify the polytypes of all the domains present in these two grains. The polytypes in G1036 include 3C and 2H (3C-2H).

The grains range in size from 650 nm (G1036) to 1000 nm (G674) to 1230 nm (G506), and in shape from triangular to highly elliptical. Crystal defects are present in all three grains in the form of high-density stacking faults, but they are limited to the 3C domains in each. Voids are present in all three grains and vary in abundance and size — >8 at 10s of nm in size and comprising ~1 area% of G1036, two large (100s of nm) and numerous small (<10 nm) and comprising ~5 area% of G506, and numerous ranging from <10 nm to 100 nm and comprising ~10 area% of G674 (see Table A1 for detailed size data). In G506, the large voids (Fig. 5b) are located at domain boundaries, whereas the smaller voids are located within domains and seem to have a random distribution that is unassociated with structural features (Fig. 5c). The voids in G674 are heterogeneously distributed (Fig. 5e). The large voids in G1036 are mostly located at domain boundaries (Fig. 5i) and the smaller voids within the 2H domains (domains are labeled in Fig. A1).

Minor elements in G506 are mostly concentrated in subgrains. The abundances of Al and N are very low (Table 2; Fig. A3) with no discernible heterogeneities in their distributions. The abundances from TEM EDX of Al and Mg in G506 are at odds with SEM-EDX analyses of this same grain by Liu et al. (2017b). Converting our data from atomic to weight percent, TEM EDX analyses show 0.2 wt.% Al and Mg below detection limit, whereas Liu et al.'s (2017b) SEM-EDX analyses show 1.1 wt.% Al and 0.6 wt.% Mg. The most likely explanation for this discrepancy is contamination, which is consistent with previous observations of Al surface contamination of presolar grains (Groopman et al. 2015, Liu et al. 2019). The FIB-section preparation method of cutting into the grain avoids surficial contamination issues, and as such, the TEM EDX data can be considered to be more accurate. Numerous small (<5–10 nm), equant, Zr-bearing subgrains are present, which also contain minor amounts of Ti, Mo, and/or V (Fig. 5c). One elongate, Ti-bearing subgrain is also present on the grain surface (labeled as subgrain 1 for G506 in Fig. A10). Some subgrains are associated with voids but most are not. Comparing EDX data from the Zr-bearing subgrains with the adjacent SiC reveals no to only slight N enrichments (as much as +1.3 at.%) and both C enrichments (as much as +3.2 at.%) and depletions (as much as -1.9 at.%). On the other hand, the elongate Ti-bearing subgrain shows a large N enrichment (+4.7 at.%) but only a slight C depletion (-1.6 at.%), implying that the subgrain is likely a TiC or TiCN. The ZrC and TiC(N) subgrains in G506 have large variations in their Zr/Ti and Mo/Ti ratios (up to 1.5 orders of magnitude) (Table 3; Fig. A7).

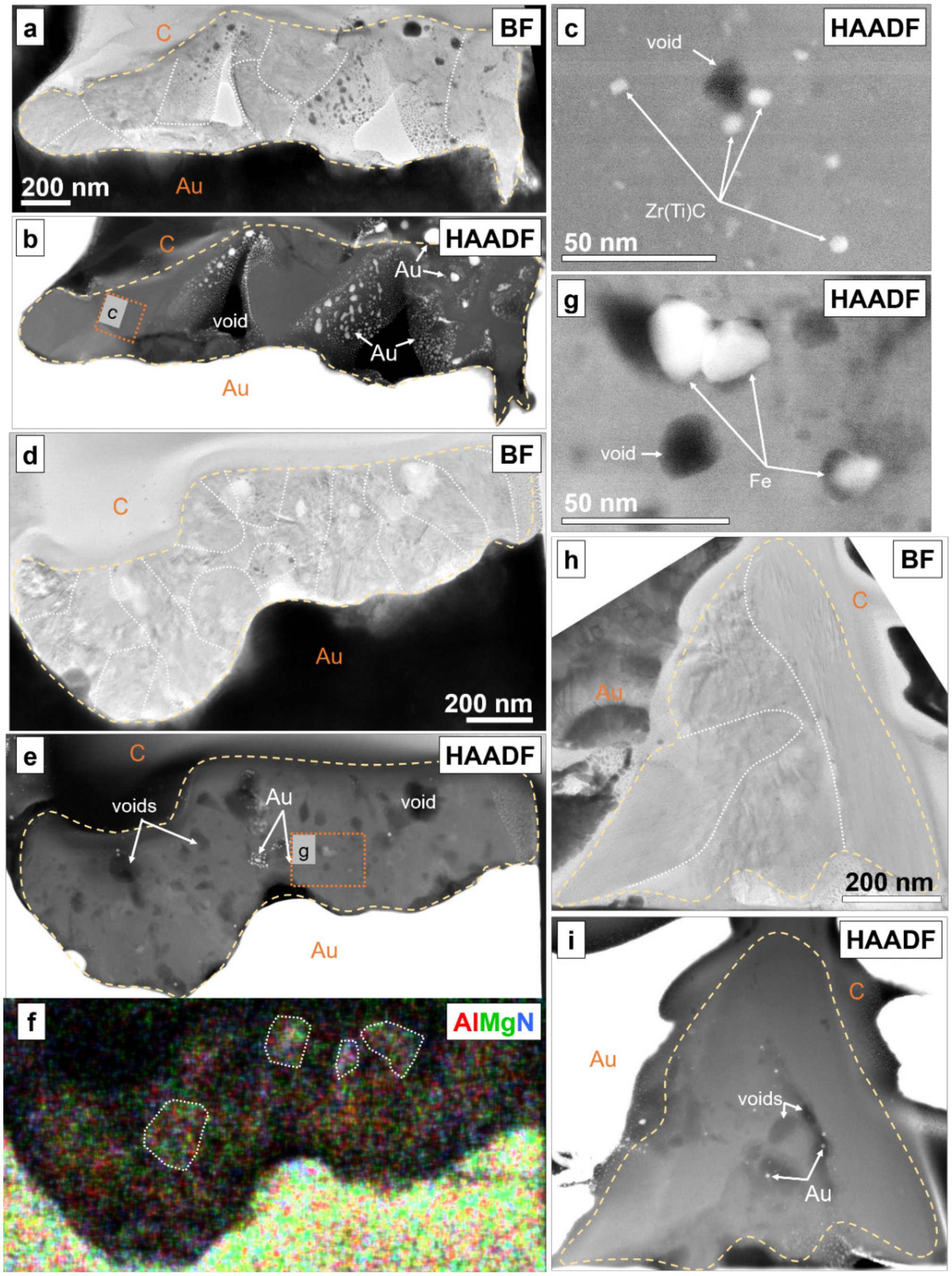

**Figure 5.** BF images, STEM HAADF images, and EDX map of X grains (a–c) G506, (d–g) G674, and (h–i) G1036. The dashed lines outline the SiC grains. The Au foil, supporting the SiC, and the protective C strap are labeled. Bright white material within the grains' HAADF images (b, e, i) is Au contamination from FIB sample preparation. All three grains are complex intergrowths with multiple crystal domains, indicated by the stippled white lines in (a, d, h). Numerous voids are present in all three grains (b, e, i). G506 and G674 contain subgrains (c, g), which are predominantly ZrC in G506 and Fe-metal/carbides/silicides in G674. The EDX map of G674 (f) shows the presence of Al,Mg,N-rich patches, a few examples of which are outlined by the stippled white lines.

In addition to Al and N, most X grains also contain Mg from the radioactive decay of $^{26}$Al to $^{26}$Mg ($t_{1/2}$ = 0.72 Myr). The EDX map of G674 (Fig. 5f) shows Al, Mg, and N spatial heterogeneities, but the Al,Mg,N-rich regions are patchy and do not appear to be associated with the grain boundary or other morphological features of the grain, unlike what is observed in MS grains G620 and G619. A few of these Al,Mg,N-rich patches are outlined in Figure 6f. Table 2 shows representative compositions of the SiC in G674 that is less enriched in Al, Mg, and N (1.9 ± 0.1 N, 0.63 ± 0.08 Mg, and 1.0 ± 0.1 Al) as well as the Al,Mg,N-rich regions (3.0 ± 0.2 N, 0.7 ± 0.1 Mg, and 1.4 ± 0.2 Al). More detailed information, including a higher magnification image/EDX map, are available in the appendix (Fig. A7). The level of crystal disorder is likely the influential factor for the distribution of the minor elements, similar to the case in G312 with the high-density stacking fault regions. G674 contains equant and elongate, Fe-bearing subgrains, some with minor amounts of Ti, Ni, and/or V, which range in size from <5 nm to 26 nm. The majority of the subgrains are associated with voids; in some cases, the subgrains appear to sit within voids themselves (Fig. 5g). Iron-bearing subgrains in X grains could be Fe metal, carbides, and/or silicides, based on theoretical models of CCSNe and previous studies of presolar SiC X grains (Hoppe et al. 2001; Lodders 2006b; Hynes et al. 2010; Gyngard et al. 2018; Kodolanyi et al. 2018). Comparing EDX data from the subgrains with the adjacent SiC reveals both Si enrichments (as much as +1.8 at.%) and depletions (as much as -3.1 at.%) as well as C depletions (varying from -0.9 at.% to -6.8 at.%). The subgrains with high Fe contents and larger depletions in Si are likely Fe metal, containing other transition metals, whereas the subgrains with Si enrichments or smaller Si depletions may be Fe silicides. For Fe carbides, such as cohenite ($Fe_3C$), a decrease in the C content would be expected, given the change in proportions of the elements from a 1:1 Si-C ratio in SiC to a 3:1 Fe-C ratio in $Fe_3C$. However, the C depletion ought to be coupled with a greater Si depletion. This is only observed in one subgrain (G674 subgrain 11 in Table 3). The subgrains in G674 have large variations in their Fe/Ti ratios (nearly two orders of magnitude) (Table 3; Fig. A7).

G1036 contains higher abundances of Al, Mg, and N compared to G674 (Table 2; Fig. A3), and there appears to be some variation in the Al, Mg, and N contents across G1036, with the bottom portions of the grain having higher minor element contents (Fig. A8; Table A2). Additionally, there does not appear to be a correlation between the minor element abundances and either the polytype (3C versus 2H) or the degree of disorder in the domains. The 3C domains in G1036 contain high-density stacking faults. Interestingly, G1036 does not contain any subgrains; the rounded, white material in Figure **5**i is Au that was redeposited during FIB-section sample preparation.

## 4. DISCUSSION

### 4.1 EDX-μ-Raman screening tool

Our TEM studies enable us to expand on the research of Liu et al. (2017b) by investigating the polytypes of eight additional presolar SiC grains with available Raman and isotope data. Table 1 lists the polytypes for each grain as determined using both methods — TEM and μ-Raman (hereafter Raman). The polytype from the Raman study is listed as either 3C (3C bulk structure) or non-3C (non-3C bulk structure), depending on the TO peak position, with 3C at 796 $cm^{-1}$, 2H at 764 $cm^{-1}$, and other polytypes with their dominant peaks between these two values (Nakashima & Harima 1997; Nakashima et al. 2013).

Figure 6 compares the Raman and TEM findings and includes both the results of the eight grains from the current study as well as those of the three grains analyzed with TEM by Liu et al. (2017b). The latter are labeled with light gray text. The TO peak position is plotted against the peak width, calculated as FWHM and shown on a logarithmic scale. As discussed in Liu et al. (2017b), their Raman spectra of the SiC showed systematic offsets in the TO peak positions of +3 $cm^{-1}$ compared with the literature values. As such, we have shifted the peak positions for the literature values of each polytype in Figure 6 by +3 $cm^{-1}$, so that the 3C reference line falls at 799 $cm^{-1}$ (796 + 3 $cm^{-1}$), the 2H reference line at 767 $cm^{-1}$ (764 + 3 $cm^{-1}$), and so forth. The symbol shape corresponds to the isotopic group of the grain, and the symbol color indicates the TEM-determined polytype. G619 is plotted twice in the figure, as the Raman spectrum for this grain contained two dominant peaks.

The data for Raman- and TEM-determined polytypes show good agreement in Figure 6, with the exception of G648 (current study) and G706 (Liu et al. 2017b). Both G648 and G706 are well-crystallized 3C SiC grains; however, the peak positions for these grains are well below the 3C reference line. For G648, we found no evidence for crystal disorder or twinning; however, the grain was lost after one TEM session, limiting the amount of data collected from it. As discussed in Liu et al. (2017b), diffraction data for G706 indicates the presence of twinning or other superstructures, although the degree of either structural feature is less than what would be expected based on the peak shift in the Raman spectrum. One possible explanation for the discrepancy between the Raman and TEM data observed for both grains is that the portion of the grain sampled by the FIB-sectioning process may not be representative of the original bulk grain as measured with Raman spectroscopy. The original grains could have contained more than one domain, at least one of which was non-3C, and/or regions of high disorder. The FIB-sectioning process could have milled away these other regions and sampled only domains that were 3C and/or with low disorder. On the other hand, this discrepancy is unlikely to have been caused by selective sampling of the grains by Raman in the Z direction, as oriented in the Au foil, given that the laser used for Raman spectroscopy (532 nm frequency-doubled Nd:YAG) has a penetration depth in SiC in the millimeter range (Harima 2006). As discussed in Liu et al. (2017b), beam damage from SIMS analyses can also be discounted as the effect of ion beam damage is negligible under the NanoSIMS analytical conditions of the study.

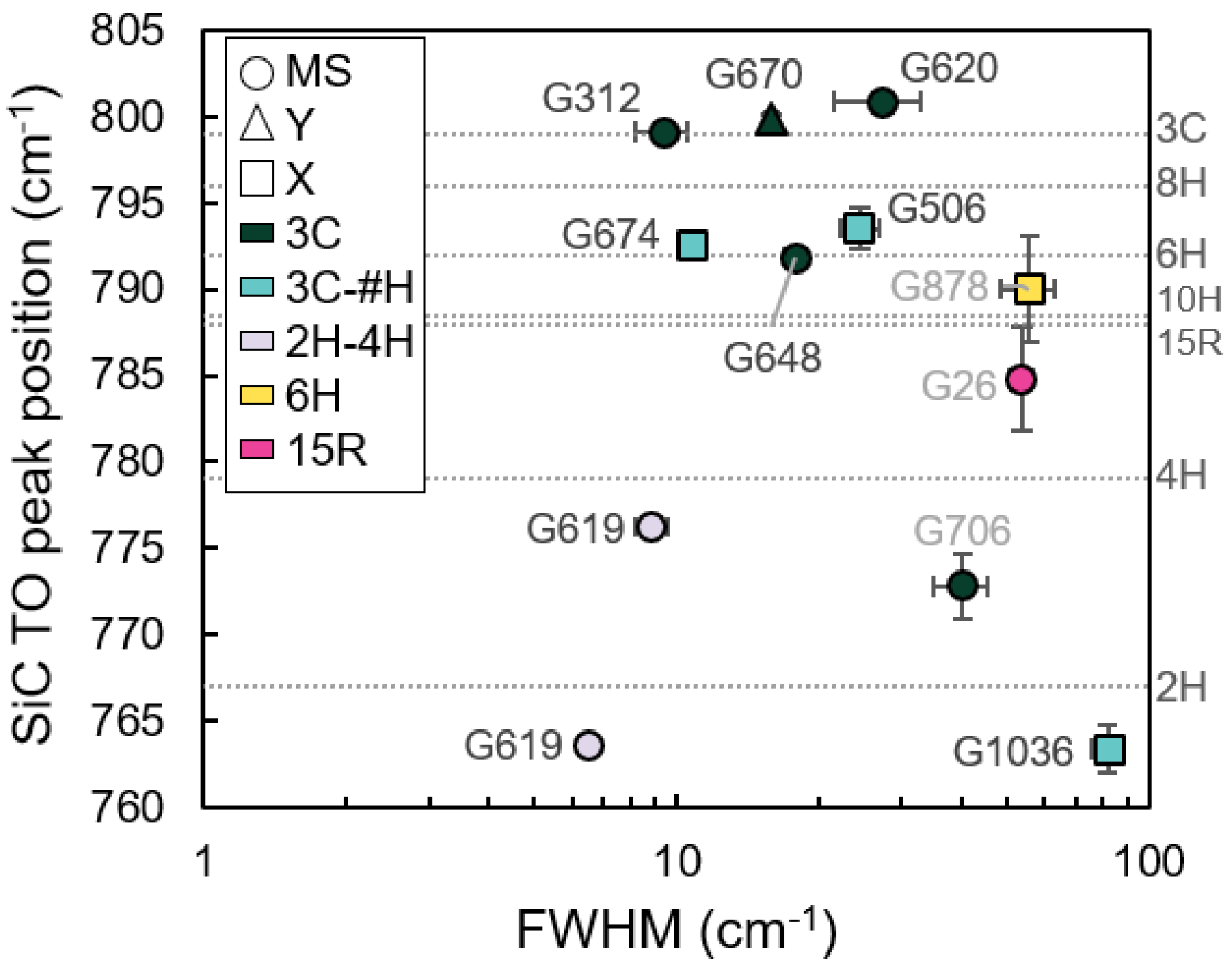

**Figure 6.** Presolar SiC grain Raman transverse optical (TO) peak position vs. peak width (measured as FWHM). The TEM-determined polytypes, indicated by the symbol color, are from the current study, with the exception of the three grains labeled with light gray text (G26, G706, G878). Those grains' TEM-determined polytypes as well as Raman spectral data for all grains are from Liu et al. (2017b). The Raman literature values for the polytypes (stippled, horizontal lines) are from Nakashima & Harima (1997) and Nakashima et al. (2013); a literature value for 14H is not available. The symbol shape indicates the isotope group. The "#" in "3C-#H" corresponds to any of the hexagonal polytypes (2H, 4H, etc.). Error bars are 1σ, some of which are smaller than the symbol size.

In contrast to these significant downshifts in TO peak position, G620 shows a slightly higher TO peak position than the 3C reference line. Prior studies have shown that N and Al impurities can cause an upshift in the peak position of the 6H polytype (Li et al. 2010; Lin et al. 2012). If the same is true of the 3C polytype, the higher impurity contents on the edges of G620 (Al,N-rich SiC) could be the cause of this upshift.

In addition to the peak position, the peak width can contain both structural and compositional information on the SiC grain. The presence of multiple domains of different polytypes, in addition to shifting the peak position, can also broaden the peak width. In general, crystal disorder, which includes structural features such as stacking faults and voids, causes peak broadening (Rohmfeld et al. 1998). For the SiC grains investigated here, broad peak widths, defined as greater than 10 $cm^{-1}$, are observed in MS grains G648 and G620, Y grain G670, and X grains G506, G674, and G1036. Of these grains, those with multiple domains, high-density stacking faults, and abundant voids include G506, G674, and G1036. G620 may have more strain than average given the presence of the large TiN subgrain and several large cracks that cross the grain along the boundary between the two crystal domains. On the other hand, G648 and G670 are 3C grains with only minor defects. The peak broadening for these grains could potentially be related to features in portions of the grains that were not sampled by the FIB section. The fact that the Raman spectra were collected from the pre-FIB-sectioned bulk grain makes it difficult to interpret the variations in peak position and peak width. Future work analyzing the FIB-sectioned grains with confocal Raman spectroscopy could provide further insight.

Although crystal disorder and intergrowths of multiple crystal domains complicate the ability of the Raman spectra to predict the exact polytypes of the SiC grains, we have demonstrated that the Raman spectra are useful as a screening tool for finding structurally unusual presolar SiC grains, i.e., those with non-3C domains and/or high degrees of crystal disorder. Consistent with the underlying mineral physics, the Raman screening tool does not appear to generate false positives; that is, if Raman spectra imply that a grain is 3C, TEM analyses confirm that it is 3C. As such, Raman spectroscopy can be a powerful, non-destructive screening tool to use as a first pass in identifying unique presolar SiC grains for further detailed TEM investigations.

### 4.2 Circumstellar environmental conditions

Isotopic compositions show that the presolar SiC grains investigated in this study condensed either in the circumstellar envelopes of AGB stars of varying mass and metallicity (MS and Y grains) or in CCSNe ejecta (X grains). Factors such as gas composition (e.g., C/O ratio), temperature, and pressure all influence which phases condense and what crystalline structures form. Thermodynamic equilibrium calculations show that certain sequences of phases condense under limited ranges of pressures and temperatures (Lodders & Fegley 1995; Sharp & Wasserburg 1995; Lodders & Fegley 1999; Ebel & Grossman 2001; Ebel 2006; Lodders 2006b). The sizes and structural features of the grains provide additional information that can be used to constrain temperatures of formation and rates of growth. By combining our observations with those of previous studies on both presolar SiC and graphite, we work towards conveying a more comprehensive understanding of both common trends and more unique characteristics of the AGB star and CCSN circumstellar environments.

#### *4.2.1 Polytypes*

Five of the eight grains we analyzed are 3C or 3C-2H, consistent with the general results for >500 grains analyzed by Daulton et al. (2003). Only three of the grains studied here contained domains with higher order hexagonal (hereafter "higher order") polytypes. The dominance of 3C and 2H polytypes is true of grains that originated in both AGB stars and CCSNe, implying that the conditions where SiC forms must be quite similar in both environments.

Temperature is a major factor influencing which polytype forms; generally, 2H is the lowest temperature polytype, with 3C forming at higher temperatures, and all other higher order polytypes forming at even more elevated temperatures (Daulton et al. 2003). The exact formation temperature varies depending on the pressure, gas composition, and the composition and structure of any substrates that act as nucleation sites for the SiC (Hynes et al. 2010 and references therein). The influence of temperature, pressure, and gas composition on when SiC is likely to condense can be approximated from thermodynamic equilibrium calculations. From such studies, equilibrium condensation temperatures for SiC are estimated to be 1390–1736 K for AGB stars (for pressures from $10^{-3}$ to $10^{-7}$ bars and C/O ratios from 1.05 to ≥2; Lodders & Fegley 1995) and 1405–1551 K for CCSNe (for a pressure of $10^{-7}$ bars and various CNO mixtures for the different zones in the parent star; Hoppe et al. 2001, Lodders 2006b).

Laboratory experiments that synthesized various SiC polytypes yielded the following formation temperatures: 2H from 1473–1723 K, 3C from 1673–2073 K, and higher order polytypes >1973

K, with each study using one growth method but the methods varying between studies (e.g., Patrick et al. 1966; Powell 1969; Berman & Ryan 1971; Bootsma et al. 1971; Krishna et al. 1971; Stan et al. 1994). However, higher order polytypes, specifically 4H and 6H, have been grown at lower temperatures (~1800 K) (Eddy et al. 2008; Myers-Ward et al. 2016). The aforementioned calculated circumstellar condensation temperatures, therefore, generally agree with the experimental temperatures, with the temperature ranges for the 2H and 3C polytypes matching most closely. The agreement suggests that the theory of thermodynamic equilibrium is sufficient in describing the circumstellar condensation environments, especially those around low-mass AGB stars, and is, therefore, the basis for the following discussion.

Based on the lower condensation temperature of 2H compared to 3C, Daulton et al. (2003) suggested that 2H grains formed at larger radii than the 3C grains in AGB star envelopes. Given that smaller radii correspond to higher pressures and higher pressures to higher densities of Si and C in the gas, it follows that the polytype that forms at smaller radii (i.e., 3C) would be the most abundant, as is observed. Using this same reasoning, higher order polytypes would not be expected to form from the perspective of thermodynamic equilibrium, given that their experimentally-determined formation temperatures are too high for regions where SiC is stable in circumstellar envelopes.

Although five of the eight studied grains contained only 3C and/or 2H domains, the remaining three grains (G619, G506, and G674) had domains that belong to the higher order polytypes. This implies that, although uncommon, there are conditions that promote the formation of higher order polytypes. One possible explanation for the formation of higher order polytypes involves thermal metamorphism, in which SiC initially formed as either 3C or 2H, but was subsequently heated and cooled rapidly enough to lock in a higher order polytype that is stable at elevated temperatures. However, this would not explain the presence of intergrown grains containing 3C and/or 2H domains (i.e., G506 and G674), as any heating ought to have thermally metamorphosed all domains rather than leaving certain domains intact. Additionally, if thermal metamorphism was responsible for the formation of higher order polytypes, then such grains would not be expected to contain subgrains that exsolved from SiC at lower temperatures. For example, the 2H-4H MS grain G619, contains the lower-temperature exsolution phase CaS (discussed in sections 4.2.2 and 4.2.3), which would not be expected in a grain reheated to higher temperatures.

Interestingly, two of the three grains with higher order polytypes are intergrowths of multiple domains and are also X grains — G506 (3C-8H) and G674 (3C-2H-10H-14H). These grains condensed in the ejecta of CCSNe and, given the inherent complexity of the CCSN environment, trying to understand their properties in terms of radial formation distance from the parent star is not as applicable as for grains from AGB stars. Additionally, the evidence for rapid growth of these two grains (discussed in section 4.2.3) likely played a role in stabilizing the formation of the higher-order polytypes.

The only presolar SiC grain identified in this study without any 3C domains is MS grain G619, which shows evidence for 2H Al,N-rich SiC overlying, at least partially, a 4H Al,N-poor SiC core. The Al,N-poor 4H core may reflect formation at high temperature. As the temperature decreased, 2H SiC formed with higher abundances of Al and N, given the similarity in the lattice structure of 2H SiC and AlN. Experimental work by Zangvil & Ruh (1988) found that AlN forms a solid solution with 2H SiC, and their diffusion couple experiments showed that AlN abundances are greater in 2H SiC compared with adjacent 4H SiC (their Table 1). These data imply that Al and N solubilities are likely greater in 2H than 4H; there is limited data on the solubility limit of Al and N in 2H, but the solubility limits of Al and N in 4H are $10^{20}$ $cm^{-3}$ and $10^{19}$–$10^{20}$ $cm^{-3}$, respectively (Linnarsson et al. 2003; Ito et al. 2015). Although presolar SiC grains with higher order polytypes are rare, they provide information on unique formation conditions in circumstellar environments. The EDX-μ-Raman screening tool is exceptionally well suited for identifying these rare grains.

#### *4.2.2 Subgrain formation by condensation versus exsolution*

Six of the seven grains we analyzed for compositional information contained subgrains. Subgrains contained inside larger presolar grains have at least two possible origins, as either gas condensates that formed prior to or concurrent with the host grain, or as solid-state precipitates formed by exsolution from the host grains as they cooled. We found evidence for subgrains that formed by either condensation or exsolution from the SiC grains studied here.

In the case of gas-phase condensation, subgrains are expected to be phases that condense at higher or similar temperatures as compared with the SiC host. If they formed at higher temperatures, these subgrains could have acted as nucleation sites for the SiC, in which case they would be roughly centrally located. Subgrains that formed before or at the same time as the SiC could also have simply stuck to a surface and been incorporated into the SiC as it grew. Finally, if the subgrains formed at similar temperatures to the SiC host, the SiC could have itself acted as a nucleation site for the subgrains (epitaxial growth). In any of these cases, these subgrains would be expected to exhibit certain features, including:

1) Equilibrium condensation temperatures equivalent to or greater than SiC;
2) Equant morphologies or shapes and crystallographic orientations otherwise not constrained by the host grain lattice;
3) Mostly random orientation distributions with respect to other subgrains present in the SiC; and
4) Higher degrees of compositional heterogeneity between subgrains within the same SiC grain (hereafter, intragrain heterogeneity), given that condensates sample a range of conditions and have different histories (e.g., T, gas phase compositions).

Presolar SiC grains with subgrains meeting most of the condensation criteria include X grains G506 and G674. The subgrains in G506 are Zr(Ti,Mo,V)C. Thermodynamic equilibrium calculations predict that ZrC, MoC, and TiC all form prior to SiC at the pressures and C/O ratios expected to form SiC (Lodders & Fegley 1995). The morphologies of the subgrains in G506 appear equant in all but one case. The subgrains appear to be randomly distributed in the SiC grain, consistent with stochastic incorporation of precondensed grains into the growing SiC lattice. For compositional information, Figure **7** illustrates the extent of intragrain heterogeneity for each grain containing subgrains, in the form of element/Ti plots normalized to the solar ratio. A large spread in the data corresponds to higher extents of intragrain heterogeneity and a small spread to intragrain homogeneity. Compositionally, the G506 subgrains show higher degrees of intragrain heterogeneity, as their Zr/Ti and Mo/Ti compositions vary by one to two orders of magnitude (Fig. **7**c).

The subgrains in G674, on the other hand, are Fe metal, carbides, and/or silicides. The exact identity of the Fe phase is not necessary for the purposes of comparing condensation temperatures given that Fe metal, carbides, and silicides are all predicted to condense at similar temperatures. In AGB circumstellar envelopes, those temperatures are nearly 300 K lower than the predicted temperature for the initiation of SiC condensation (Lodders & Fegley 1995). However, in CCSNe, the appropriate environment for X grain G674, Fe metal and silicides may be expected to form at higher temperatures than SiC. For example, for a specific mixture of the He/C and Fe/Ni supernova zones, Lodders (2006b) found metal and silicide condensation temperatures at $10^{-7}$ bars of 1434 K and 1528 K, respectively, compared to 1405 K for SiC. Most of the subgrains in G674 appear equant, although a few are elongate. For the few subgrains imaged at high resolution, crystallographic planes did not appear to be shared between the subgrains and the host SiC. Compositionally, the G674 subgrains exhibit higher degrees of intragrain heterogeneity, given the one to two orders of magnitude variation in Ni/Ti and Fe/Ti compositions (Fig. 7f).

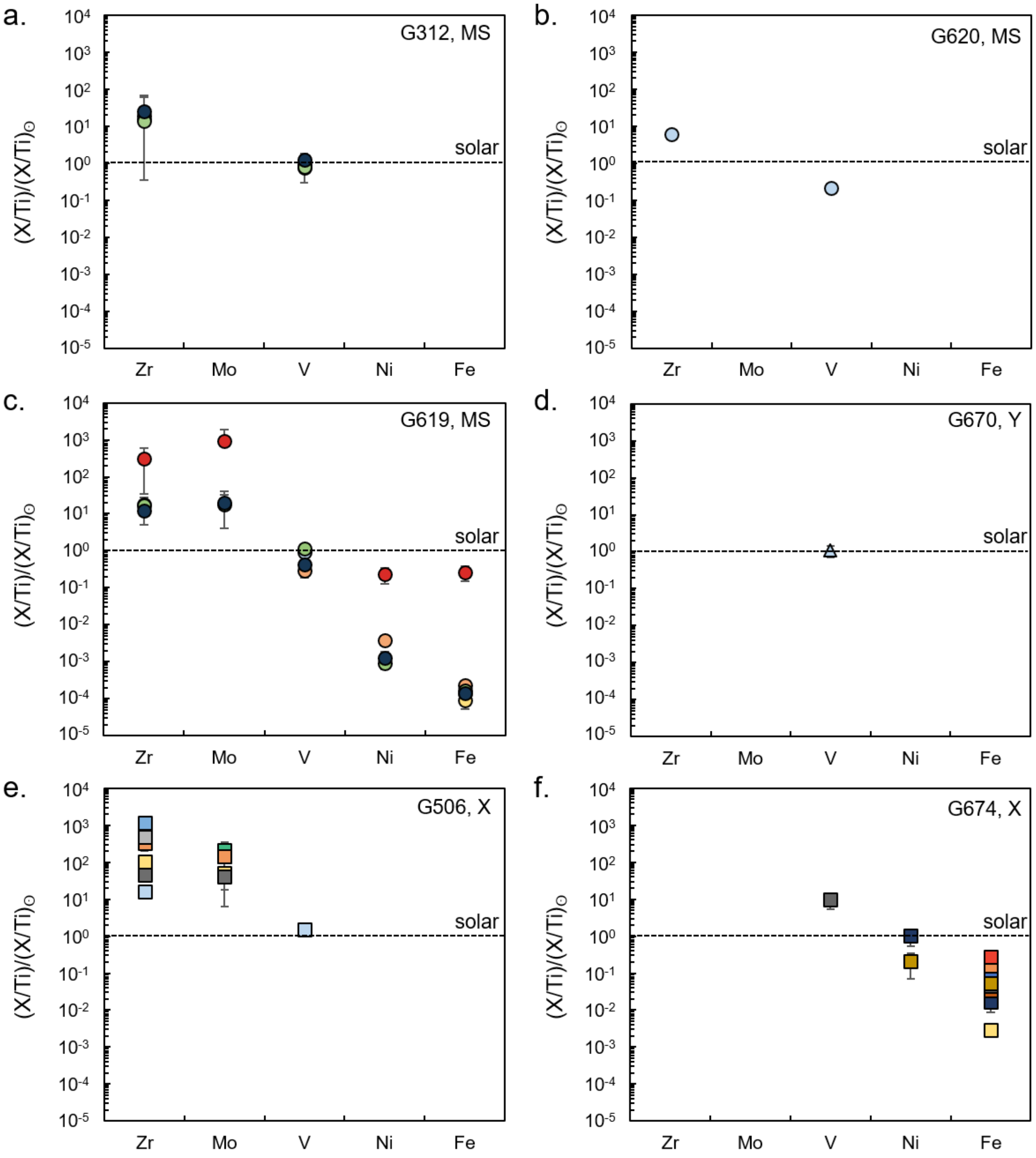

**Figure 7.** Compositional variation of subgrains in the presolar SiC grains studied. Each plot is for a different grain — (a) MS 312, (b) MS G620, (c) MS G619, (d) Y G670, (e) X G506, and (f) X G674. Only subgrains containing Zr, Mo, V, Ni, and/or Fe > 0.1 at.% and Ti above detection limit are shown. The data are presented as ratios against Ti and normalized to the solar ratio (dashed horizontal line, with solar values from Lodders 2003). Elements on the x-axis are listed in order of decreasing condensation temperature for a C/O ratio of 1.05 and pressure of $10^{-5}$ bars (Lodders & Fegley 1995). The symbol color is the same within a given plot for the same subgrain. Error bars are 1σ.

The MS grains G620 and G619 also contain subgrains with some characteristics consistent with formation by condensation. G620 only contains one Ti(Zr,V)N subgrain; therefore, the degree of intragrain heterogeneity and the distribution of subgrains within the SiC grain could not be established. Nevertheless, at C/O ratios ≤ 0.98, TiN forms at higher temperatures than SiC (Ebel 2006). Additionally, the Ti(Zr,V)N subgrain is subhedral, equant, and large. The presence of one large subgrain is not consistent with exsolution, where it would be more energetically favorable for multiple smaller subgrains to form. G619, on the other hand, contains numerous subgrains. As will be discussed below, the majority of these have characteristics consistent with formation by exsolution, but subgrain 3 is an exception, given that it is a Fe metal grain, is equant, and has a distinct composition from the other subgrains present (red data point in Fig. 7cd).

In the case of formation by exsolution, the subgrains are expected to be lower temperature phases that, on slow cooling, exsolved out from the SiC. These subgrains would be expected to exhibit certain, including:

1) Formation temperatures equivalent to or lower than SiC (for phases that exsolve from SiC, their condensation and exsolution temperatures are less than that of SiC);
2) Elongate or equant morphologies and crystallographic orientations constrained by the host grain lattice;

3) Preferential crystallographic orientations with respect to other subgrains present in the SiC; and
4) Lower degrees of compositional heterogeneity between subgrains within the same SiC grain (hereafter, intragrain homogeneity).

Presolar SiC grains with subgrains meeting the exsolution criteria include MS grains G312 and G619 and Y grain G670. G312 contains two AlN subgrains and numerous Ti(V,Zr)C subgrains, which predominantly occur in the high-density stacking fault regions. At present, thermodynamic equilibrium condensation calculations for C-rich stellar environments do not consider solid solutions, but they predict that pure AlN forms at temperatures below that for the start of pure SiC condensation, whereas pure TiC forms at higher temperatures than SiC (e.g., AlN = 1242 K, TiC = 1702 K, SiC = 1544 K for $10^{-5}$ bars and a C/O ratio of 1.05; Lodders & Fegley 1995). However, all but one of the TiC subgrains contain N enrichments (0.6–1.5 at.% greater N relative to the adjacent SiC), which potentially indicates formation at lower temperatures than pure TiC. Additionally, the morphologies of the TiC subgrains are mostly elongate and show a preferred orientation with the direction of elongation parallel to the stacking faults (Fig. 2b). The TiC subgrains in G312 vary by less than one order of magnitude, indicating intragrain homogeneity for Zr/Ti and V/Ti (Fig. 7a).

G619 contains one CaS subgrain and numerous Ti(Zr,V,Fe,Ni,Mo)C subgrains, which predominantly occur in the Al,N-rich SiC. Thermodynamic equilibrium calculations predict that CaS forms at temperatures below the initiation of SiC condensation (CaS = 1220 K versus SiC = 1544 K for $10^{-5}$ bars and a C/O ratio of 1.05; Lodders & Fegley 1995). Again, pure TiC is predicted to form at higher temperatures than SiC. As with G312, the morphologies of the TiC subgrains are mostly elongate and show a preferred orientation, but, in this case, with the direction of elongation perpendicular to the Al,N-rich bands that outline the euhedral, Al,N-poor SiC core (Figs. 3b, 3d). The CaS subgrain is located adjacent to a void and a TiC subgrain, with which it appears to share a sharp crystal boundary and crystallographic orientation relationships based on HRTEM images (Fig. A5). The void may have acted as a nucleation site for both as they exsolved from SiC. Similar to G312, the TiC subgrains in G619 show intragrain homogeneity and may also contain N enrichments. Excluding the Fe metal subgrain (subgrain 3), which is likely a pre-exiting condensate that was incorporate into the SiC as discussed above, the Zr/Ti, Mo/Ti, Ni/Ti, and Fe/Ti compositions vary by less than one order of magnitude (Fig. 7c).

The subgrains in G670 are TiC with only minor V in one subgrain. As with G312 and G619, the TiC subgrains in G670 have mostly elongate morphologies, but do not show preferred orientations with respect to one another or any other structural or compositional features in the grain. Unfortunately, given the lack of elements other than Ti in the G670 subgrains, we were not able to establish whether there was intragrain heterogeneity (Fig. 7d). The elongate morphologies are the strongest evidence for the formation of the TiC subgrains by exsolution. Intragrain homogeneity and preferred orientations of subgrains with respect to one another and other structural and/or compositional features thus argue for the formation of the TiC subgrains in G312, G619, and G670 by exsolution rather than condensation. This is at odds with most interpretations of TiC subgrains in graphite, in which they are often postulated to have formed by condensation (e.g., Bernatowicz et al. 1996; Croat et al. 2003). However, it is important to note that, although TiC initially forms at high temperature, its stability field extends to lower temperatures as well, so that it can form concurrently with SiC. Additionally, although elongate presolar grains that formed by condensation have been observed previously (needle-like SiC grains in Daulton et al. 2003; Liu et al. 2017b), it is difficult to envision a process whereby several elongate TiC condensates were all incorporated into the same SiC grain, especially one with a low abundance of subgrains (i.e., G670).

An alternative formation mechanism for these TiC subgrains is that they condensed on the SiC. This explanation is most appropriate for G619, in which exsolution is difficult to reconcile with some textural observations. Specifically, if the TiC subgrains formed by exsolution and required enough time for Ti to diffuse through the SiC to the precipitates, the Al,N zonation may have not been preserved, depending on the relative diffusion rates of Al, N, and Ti in SiC. Unfortunately, given the lack of diffusion rates of these three elements in SiC, it is impossible to reach a more definitive statement about the likelihood of G619's TiC subgrains having formed by exsolution or co-condensation. In any case, formation of the TiC subgrains in presolar SiC by exsolution or co-condensation is also consistent with the observations of TiC subgrains in SiC by Bernatowicz et al. (1992).

### *4.2.3 Formation mechanisms and conditions*

With our understanding of which subgrain populations in each SiC grain likely formed by condensation and which by exsolution, we can better constrain the formation histories of the presolar SiC grains as well as the conditions required for the structures and compositions observed. In the following section, we will discuss the likely formation scenarios and environmental conditions for the seven presolar SiC grains for which we have compositional data. We can use models developed by Lodders & Fegley (1995) and Ebel (2006) to predict the C/O ratios under which the grains formed. These models are best applied to the MS grains, given that the models are for solar metallicity under thermodynamic equilibrium; however, such models can also be used for Y grains, since lowering the metallicity of an AGB star only serves to decrease the condensation temperatures without changing the relative condensation sequence of the phases, with the exception of graphite (Ebel 2000; Lodders 2006a). These same models can also be applied to X grains, with the caveat that formation conditions in CCSN outflows are more complex than those in AGB envelopes, involving mixing between zones and more rapid crystallization (Hynes et al. 2010; Liu et al. 2017b; Gyngard et al. 2018; Kodolanyi et al. 2018). In any case, thermodynamic equilibrium calculations represent a useful tool for interpreting the history of X grains.

The crystallization paths of the MS grains in temperature-C/O ratio space are shown in Figure 8a. The SiC stability field in this plot includes thermodynamic data for both α-SiC (all non-3C polytypes) and β-SiC (3C) (D. Ebel, personal communication, 2020 June 4). The C/O ratio for the gas is not expected to change over the timespan during which a SiC grain forms (on the order of one year for AGB stars, from Sharp & Wasserburg 1995), hence the vertical crystallization paths. The stability fields indicate at what temperatures and C/O ratios a given phase is stable under equilibrium conditions. However, not all phases need be observed down a particular crystallization path. Other considerations, such as the abundance of the elements in the gas and kinetics, determine whether a phase will indeed form.

*MS grain G312*

G312 contains regions of both low- and high-disorder SiC. The latter contains AlN subgrains and numerous TiC subgrains, which appear to be oriented parallel to the stacking faults. Given that the AlN and TiC subgrains have characteristics most consistent with formation by exsolution, the formation sequence for the different components in G312 is as follows: Low-disorder SiC → high-disorder SiC → TiC subgrains → AlN subgrains. Changes in temperature and pressure as the SiC grain formed could have caused the solubility limits of the minor elements in SiC to increase or caused condensation to occur rapidly enough that impurities were trapped in the SiC.

A high-impurity content could have promoted the formation of the high-density stacking faults that characterize the high-disorder SiC. Lower temperatures correspond to higher sticking coefficients for Al, N, and Ti, as well as Si and C. Depending on the solubility limits of each of these elements in SiC, a change in the ratio of these sticking coefficients (Al, N, Ti versus Si and C) can have an influence on the incorporation of the minor elements upon cooling. On further cooling, those higher abundances of impurities would have then formed discrete phases via exsolution. Given that the Gibbs free energy of formation for TiN is more negative (implying that it is more stable) than that of TiC at temperatures <1800 K, we might expect that TiC with greater N abundances may exsolve at lower temperatures than pure TiC (Chase, 1998). The elevated N content (0.6–1.5 at.% greater than the adjacent SiC) in all but one of the TiC subgrains then implies that these grains formed at lower temperature than pure TiC, which is consistent with the formation sequence proposed. The presence of AlN subgrains in G312 implies that the grain cooled down to low temperatures; the formation temperature of AlN in solid solution with SiC is only slightly higher than the condensation temperature (a difference of 32 K at $10^{-5}$ bars and a C/O ratio of 1.05 from Lodders & Fegley 1995).

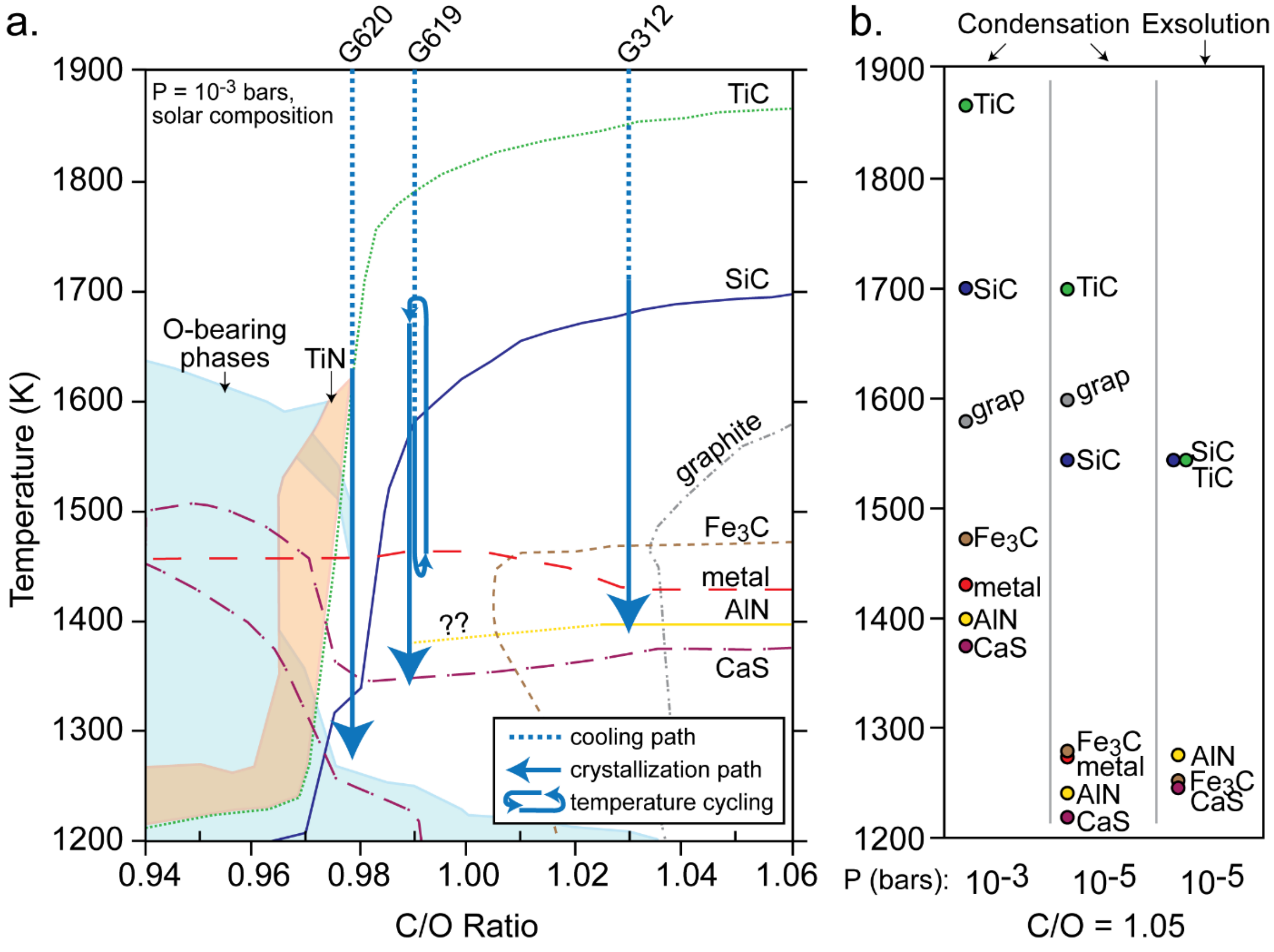

**Figure 8.** Formation temperatures of circumstellar phases. (a) Equilibrium condensation plot for a solar composition gas as a function of C/O ratio and temperature. The lines, sourced from plate 9 in Ebel (2006), represent stability fields (50% condensation temperatures) for the phases labeled and are for a pressure of $10^{-3}$ bars. The blue arrows represent crystallization paths for the formation of MS grains G312, G620, and G619. (b) Comparison of formation temperatures of phases at a C/O ratio of 1.05 as a function of pressure ($10^{-3}$ vs. $10^{-5}$ bars) and formation mechanism (condensation vs. exsolution from SiC). The data points are sourced from Ebel (2006) (far left) and from Lodders & Fegley (1995) (center and far right). A decrease in pressure corresponds to a decrease in condensation temperature, with the exception of graphite. For phases with condensation temperatures less than SiC, the exsolution temperatures are roughly similar to the condensation temperatures.

Figure 8a illustrates one possible formation path for G312 at a C/O ratio of 1.03; however, any C/O ratio ≥1.03 is also consistent with the characteristics of this grain. At C/O ratios less than about 1.03, information on the stability of AlN is not available in the current literature (i.e., Ebel 2006). However, if AlN is indeed stable at lower C/O ratios, G312 could also have formed at any C/O ratio >0.98. The different stages in the formation history of G312 are summarized in Figure 9a.

*MS grain G620*

G620 contains Al,N-poor and Al,N-rich SiC, as well as a large TiN subgrain at the boundary between these two types of SiC. We can exclude the possibility that the Al,N-rich SiC is the result of contamination that is sometimes observed in NanoSIMS ion images (e.g., Groopman et al. 2015; Liu et al. 2019) since the Al contents in these regions of G620 are at much greater concentrations than what is observed in contaminated materials. Additionally, these regions are also enriched in N in G620, and an association between Al contamination and N has not been noted. As discussed previously, the TiN subgrain has characteristics that are most consistent with formation by condensation and incorporation into SiC. As the Al and N zonation shows, G620 likely represents SiC that condensed in stages. The euhedral SiC core represents SiC that condensed at higher temperatures. At much lower temperatures when CaS began to condense, some of the Si in the gas phase, previously bound to SiS, was freed and able to form additional SiC (Sharp & Wasserburg 1995). This resulted in a second stage of SiC condensation at much lower temperatures, more akin to the condensation temperatures of AlN. Hence, the formation sequence for the different components in G620 is: TiN subgrain → Al,N-poor SiC → Al,N-rich SiC (during/after CaS condensation). Figure 8a illustrates a possible crystallization path for

G620 as a function of C/O ratio. The path at a C/O ratio of 0.98 is the highest C/O ratio that crosses

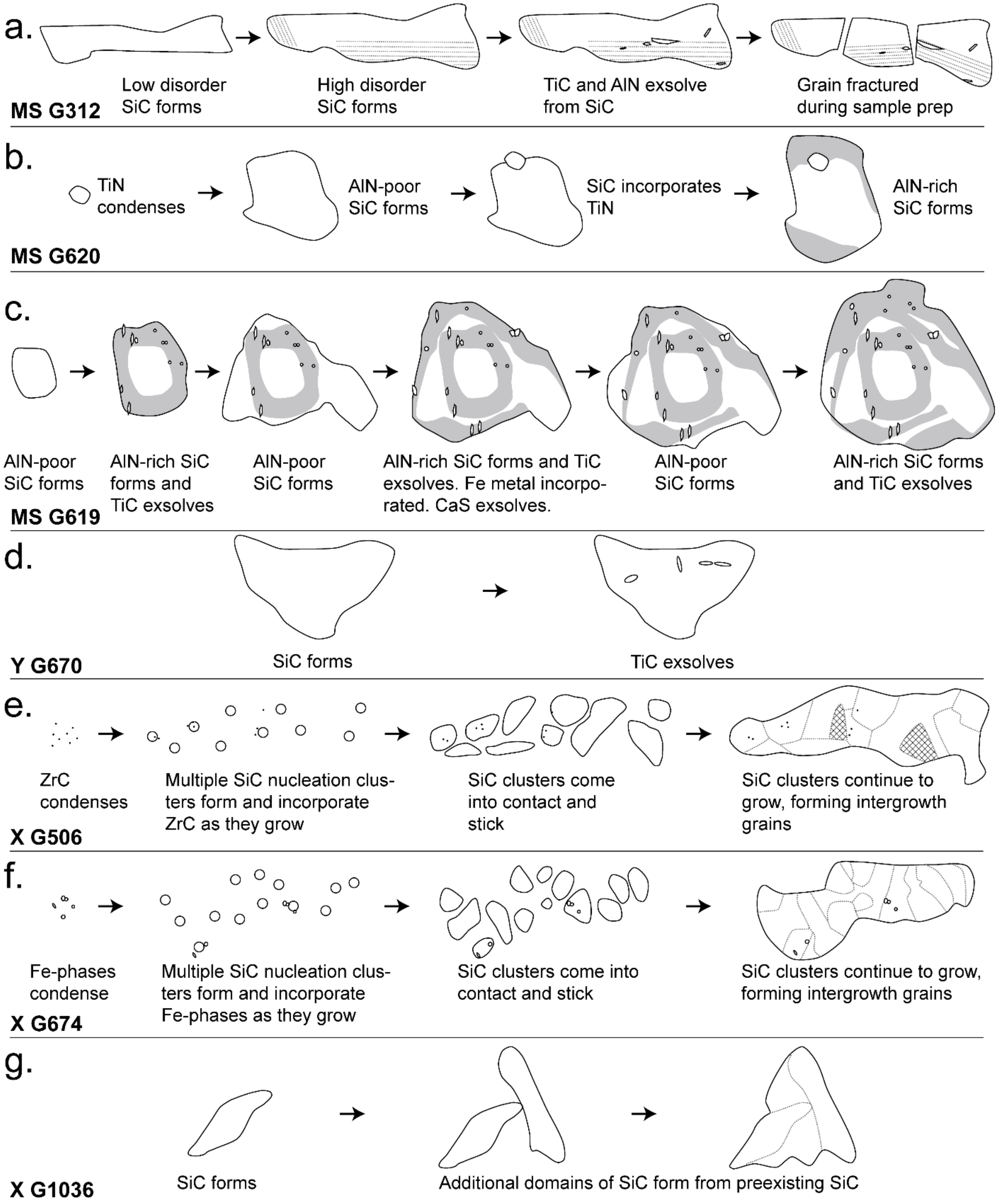

**Figure 9.** Formation history of MS grains (a) G312, (b) G620, (c) G619, Y grain (d) G670, and X grains (e) G506, (f) G674, and (g) G1036. Structural and compositional features of the grains provide evidence

for rapid cooling, high impurities in the circumstellar envelope gas, high dust-to-gas ratios, temperature cycling, and/or changes in pressure as the SiC grains formed. See text for discussion.

the TiN stability field. At lower C/O ratios, oxides and silicates (O-bearing phases) are predicted to be stable, which is inconsistent with the phases we observed. Given the presence of Zr and V in the subgrain as well as the potential that it is a carbo-nitride rather than a pure nitride, the formation path may lie on the more C-rich side of the TiN stability field, in which case, SiC would condense prior to CaS. Figure 9b summarizes the different stages in the formation history of G620.

*MS grain G619*

G619 contains at least two generations, possibly three, of Al,N-poor and Al,N-rich SiC as well as TiC subgrains, an Fe-bearing subgrain, and a CaS subgrain in association with the Al,N-rich SiC. As discussed previously, the TiC and CaS subgrains have characteristics that are most consistent with formation by exsolution, whereas the Fe-bearing subgrain was probably precondensed and incorporate into the SiC during its growth. The complex Al and N zonation shows that the SiC in G619 likely condensed in multiple stages. The Al,N-poor SiC core represents SiC that condensed at the highest temperatures. As with G620, the Al,N-rich SiC likely formed at lower temperatures. Of the subgrains that formed by exsolution, CaS represents a lower temperature phase and TiC represents a higher temperature phase for C/O ratios greater than ~0.975 (Fig. 8a), although exsolution temperatures need not be as elevated as condensation temperatures (Fig. 8b). The formation sequence for the different components in G619, based on observations from the grain itself, is as follows: Al,N-poor SiC → Al,N-rich SiC → TiC → Al,N-poor SiC → Al,N-rich SiC → TiC → Fe-bearing (precondensed and incorporated) → CaS → Al,N-poor SiC → Al,N-rich SiC → TiC. This sequence implies that G619 experienced temperature cycling, changes in pressure, and/or changes in gas composition during its growth. Figure 8a illustrates a possible crystallization path for G619 as a function of C/O ratio. The path at a C/O ratio of 0.99 has a smaller temperature difference between SiC and lower temperature phases (e.g., AlN as a proxy for Al,N-rich SiC, CaS). As compared with higher C/O ratios, this path requires less extreme changes in temperature for the temperature cycling. Changes in pressure resulting from dust condensation could also be responsible for the textural and compositional features present in G619. The effects of changes in pressure are illustrated in Figure 8b. Comparing the formation temperatures of the phases on the far left ($10^{-3}$ bars) and center ($10^{-5}$ bars) shows that a decrease in pressure roughly corresponds to a decrease in the formation temperature of the phases, although it is not a one-to-one correlation, which is an alternative explanation for the formation sequence of the grain. Figure 9c summarizes the different stages in the formation history of G619.

If changes in temperature are responsible for the features observed in G619, the lower-temperature phases contain information on the AGB circumstellar environment at lower temperature. This is significant, given the tendency for lower-temperature phases to more readily alter in meteorite hosts and/or during sample processing. The presence of these phases enclosed in SiC, which is much more resistant to alteration, provides a rare opportunity to study lower-temperature circumstellar phases.

*Y grain G670*

G670 is the only Y grain investigated in this study and, as such, is thought to come from a lower-metallicity and/or higher-mass AGB star compared to the parent stars of the MS grains. Such stars are predicted to more efficiently dredge up material from their interiors and, consequently, have higher concentrations of *s*-process elements. Limited TEM analyses have been performed on Y grains; however, isotopic and trace element compositional work by Amari et al. (2001) observed higher concentrations of *s*-process elements such as Ti, V, Y, Zr, Ba, and Ce, relative to the average MS grains.

Interestingly, the characteristics of the TiC subgrains imply that they may have formed by exsolution from the SiC rather than condensation. The formation sequence for the different components in G670 would then be SiC → TiC. Using thermodynamic equilibrium calculations for low-metallicity AGB stars (Ebel 2000; Lodders 2006a) in conjunction with the obervation of SiC and absence of oxygen-bearing phases, the relevant C/O ratio for G670 formation (>0.98) is loosely constrained. We, therefore, exclude G670 from Figure 8a, since the presence of exsolved TiC subgrains cannot narrow down the inferred C/O range. Figure 9d summarizes the different stages in the formation history of G670. The current investigation is the first TEM study to analyze the composition of a Y grain and only the second to investigate a Y grain's structural features (the other study being that of Daulton et al. 2006). Additional TEM studies of these rare presolar SiC grains are warranted and could shed more light on unusual AGB stars.

*X grain G506*

G506 is one of the three X grains studied and, as such, condensed in the ejecta of a CCSN. G506 contains (Zr,Ti)C subgrains with characteristics that are consistent with condensation and subsequent incorporation into the SiC host. The high concentration of these subgrains in some domains indicates that the SiC grain formed at high dust-to-gas ratios. The formation sequence for the different components in G506 is as follows: (Zr,Ti)C → SiC. The complex intergrowth structure, presence of large voids, and presence of stacking faults in some of the 3C domains in G506 implies the grain cooled quite rapidly. Rapid cooling promotes the formation of crystal defects, which itself results in intergrowths of multiple polytypes and stacking faults. Rapid cooling is also more likely to form voids as growth planes from multiple crystal domains impinge on one another and are no longer in contact with the gas. These observations are consistent with previous TEM studies of CCSN-sourced grains, which provide evidence for dust formation under conditions of supersaturation and/or rapid crystallization (Hynes et al. 2010; Liu et al. 2017b; Gyngard et al. 2018; Kodolanyi et al. 2018).

Although ZrC is predicted to condense at higher temperature than TiC, the lack of phase fields for ZrC in Fig. 8a (and in the source data from Ebel 2006) at varying C/O ratios makes it difficult to estimate probable C/O ratios for the formation of G506. Additionally, we exclude G506 from Figure 8a given that the differences in pressures and gas compositions between CCSNe ejecta and AGB envelopes makes a comparison of the crystallization paths of the MS and X grains misleading. Figure 9e summarizes one possible scenario for the formation history of G506, where high supersaturation of the gas in Si and C allowed for numerous nucleation clusters of SiC to form. These clusters eventually stuck to one another and continued to grow, forming intergrowths of multiple crystal domains. Alternatively, the different domains could have formed off of one another with preexisting domains serving as substrates for subsequently formed domains of SiC, although the lack of orientation relationships between adjacent SiC domains makes this explanation less satisfactory.

*X grain G674*

G674 contains Fe-bearing subgrains with characteristics consistent with formation by condensation before or at the same time as SiC. Although dust formation around CCSNe is more complex than around AGB stars, there is evidence that condensation can occur under near equilibrium conditions (e.g., Messenger et al. 2005). Assuming such conditions apply, the formation of Fe-bearing phases prior to SiC requires an Fe enrichment in the gas, relative to solar (Croat et al. 2003), e.g., from mixing of Fe-rich material from deep in the CCSNe into the He/C zone where SiC and graphite most likely form (Lodders 2006b). The isotopic compositions of the Fe-bearing subgrains have the potential to provide further detail on mixing between zones during the explosion, but are beyond the scope of this work. Using the condensation behaviors of the phases from Figure 8a as a proxy for what might be expected for dust formed in CCSN ejecta, C/O ratios ≥1.01 are most appropriate for G674. At lower C/O ratios, the condensation temperature of Fe-carbide decreases dramatically, making condensation of the Fe-carbide prior to SiC even less likely. Given differences in the pressures and gas compositions between CCSN ejecta and AGB star envelopes, we have not included G674 in Figure 8a.

In addition to the complex history indicated by the presence of Fe-bearing subgrains that condensed prior to SiC, G674, like G506, has several characteristics consistent with rapid crystallization. These include intergrowths of multiple crystal domains of varying polytypes, variable degrees of crystal disorder (specifically stacking faults in the 3C domains), minor element (Al, Mg, N) heterogeneities, abundant voids, and a high abundance of subgrains. Although much smaller in size, the complexity observed in G674 is consistent with that found in the extremely large (~25 µm) X grain Bonanza (Gyngard et al. 2018). Similar to G506, the structural complexity of G674's multiple crystal domains could be the result of either several initial nucleation sites that impinged on one another as they grew, eventually forming a single grain, or one initial nucleation site that served as a substrate for subsequently formed crystal domains. Figure 9f summarizes the different stages in the formation history of G674.

*X grain G1036*

G1036, again from a CCSN, has very high minor element abundances (Al, Mg, N) but does not contain any subgrains. Assuming equilibrium condensation conditions, any C/O ratio ≥0.98 is consistent with our observations of G1036, given the lack of subgrains to help constrain the ratio. As with G506 and G674, G1036 is excluded from Figure 8a. G1036 has several characteristics that are consistent with rapid crystallization, which include intergrowths of multiple crystal domains of varying polytypes, crystal disorder (specifically stacking faults in the 3C domains), and abundant voids. Similar to the other two X grains, G1036 likely formed from either several initial nucleation sites that impinged on one another as they grew or one initial nucleation site that served as a substrate for subsequently formed crystal domains. The latter seems most probable for G1036, given the larger size and smaller number of domains and the absence of subgrains, compared with G506 and G674. Figure 9g summarizes the different stages in the formation history of G1036.

### *4.2.4 Comparing subgrains in presolar SiC and graphite*

Both presolar SiC and graphite are purported to form in AGB circumstellar envelopes (although graphite likely formed in lower-metallicity stars; Amari et al. 2014) and in CCSN ejecta, with both phases requiring reducing conditions (C/O ratio ~≥1). Despite these broad similarities, these two populations of presolar grains have fundamental textural and compositional differences, reflecting their distinct formation conditions. In particular, we can gain considerable insight from assessing the similarities and differences between the subgrains present in presolar SiC and graphite grains by combining the findings from this study with those of previous studies on both presolar SiC and graphite grains.

A general observation from this study is that subgrains from SiC grains which formed in CCSN ejecta (X grains) tend to be condensates, whereas those from SiC grains which formed in AGB star envelopes (MS and Y grains) tend to be the products of exsolution. However, this trend does not necessarily apply to previous studies. Given the limited dataset currently available on subgrains, more TEM analyses of presolar SiC grains from these different circumstellar environments are required.

For dust that formed around AGB stars, prior investigations have noted TiC subgrains in both presolar graphite and SiC (Bernatowicz et al. 1992; Bernatowicz et al. 1996; Liu et al. 2017b). Similarly, this study has identified TiC in MS grains G312 and G619 and Y grain G670; however, it is interesting to note that these TiC subgrains are likely the products of exsolution or co-condensation (low-temperature condensates). High-temperature TiC condensates were notably absent from the grains we studied that formed around AGB stars, in contrast to observations of TiC subgrains in some graphite presolar grains. Although TiC ought to have condensed at higher temperatures and been incorporated into SiC, low Ti contents in the gas may have precluded its ability to condense. Instead, Ti may have been incorporated into SiC in trace amounts and formed TiC from exsolution or else used SiC as a nucleation site for co-condensation. This study is one of the few to identify relatively low-temperature phases (e.g., AlN, CaS) from AGB stars. These phases have not been observed in presolar graphite to date. If lower temperature phases are not found in graphite, the relatively higher formation temperatures for graphite (e.g., 1600 K) as compared with SiC (e.g., 1544 K down to 1000 K) is likely to be the cause (temperatures for $10^{-5}$ bars and a C/O ratio of 1.05; Lodders & Fegley 1995). Alternatively, the crystal structure of SiC may better accommodate these impurities as compared to graphite, which is not expected to be able to incorporate minor elements.

For dust that condensed around CCSNe, prior investigations have noted TiC subgrains in both presolar graphite and SiC (Croat et al. 2003; Hynes et al. 2010; Gyngard et al. 2018; Kodolanyi et al. 2018). This study identified TiC in one of the three X grains analyzed (G506), and also was the first to identify ZrC in dust that formed around CCSNe. Bernatowicz et al. (1996) observed Zr,Mo carbides as well as Fe-, Ni-, and Ru-bearing subgrains in graphite grains. However, that study did not determine the isotopic compositions, and hence the circumstellar sources, of the graphite grains. A subsequent study by Croat et al. (2003), which focused solely on presolar graphite grains from CCSNe, did not identify Zr,Mo carbides, however.

Iron-bearing subgrains (metals, silicides, and/or carbides) were found to be common in X grain G674 and have been observed in other SiC X grains where they were classified as either silicides (Hynes et al. 2010; Kodolanyi et al. 2018), metal (Kodolanyi et al. 2018), or undetermined Fe-bearing phases (Liu et al. 2017b; Gyngard et al. 2018). Iron metal was also noted in CCSN presolar graphite grains attached to TiC subgrains and, more rarely, as solitary subgrains (Croat et al. 2003). As mentioned previously, Fe-bearing phases (whether metal, silicides, or carbides) typically have lower condensation temperatures than either graphite or SiC (Lodders & Fegley 1995). Incorporation of Fe-bearing phases into graphite or SiC was previously explained by the following scenarios: (1) the Fe subgrains co-condensed in solid solution with SiC and later exsolved (Hynes et al. 2010), (2) they condensed prior to graphite and/or SiC, or (3) thermodynamic equilibrium was not maintained in CCSNe in the regions in which these subgrains formed (Ebel & Grossman 2001). None of these scenarios, however, is satisfying in accounting for the observations in detail. Scenario 1 does not explain the compositional intragrain heterogeneity of Fe-bearing subgrains in G674. Scenario 2 was invoked for G674 and, again, requires Fe enrichment in the gas, relative to solar, as would be expected by mixing of the He/C and Fe/Ni zones in CCSNe. In the case of mixing, Scenario 2 only holds for SiC, as graphite is still predicted to condense at higher temperatures in the He/C zone than Fe-bearing phases in the Fe/Ni zone. Scenario 3 is a possibility, but as mentioned for G674, there is evidence that thermodynamic equilibrium can be maintained in CCSN outflows. Whichever of these scenarios is the most accurate, the Fe-bearing subgrains are present in both SiC and graphite grains from CCSNe and seem to be relatively common.

In this study, we identified phases not previously observed (ZrC) or only rarely observed (CaS, TiN, and discrete AlN grains) in presolar SiC grains. TiN and AlN have been noted in X grains (Liu et al. 2017b; Gyngard et al. 2018), and CaS has been previously identified only in AB grains (Hynes 2010). This study represents the first observation of CaS in a MS grain. The ability to identify previously unobserved or rarely observed phases is largely possible owing to the unique capabilities of the aberration-corrected STEM (allowing for observation of ZrC) with a windowless EDX detector (allowing for measurement of N) as well as sample preparation using FIB sectioning (preserving CaS) rather than ultramicrotoming (introducing water which destroys CaS). With continued advances in instrumentation, TEM analyses on presolar SiC grains chosen for study using the EDX-μ-Raman screening tool may help to identify condensates not yet observed, or, if not found, help identify errors in thermodynamic equilibrium calculations or the thermodynamic data that underlie them.

**4.3 Implications for IR spectroscopy of circumstellar dust**

IR spectroscopy is used in astronomical observations of circumstellar environments to identify molecules and mineral phases by the presence of emission and absorption features, mostly in the ~0.75–300 μm region of the electromagnetic spectrum. Given that both IR and Raman spectroscopy can be used to elucidate information on the vibrational modes of materials, our observations of the influence of structural and compositional features in presolar SiC grains on their Raman spectra implies that these same features may also influence the IR spectra. Specifically, higher order (non-3C,2H) polytypes, intergrowths of multiple polytypes, crystal defects, voids, impurities (e.g., Al and N), and subgrains could, in principle, all affect IR spectra to the extent that grains with these features are missed entirely or misinterpreted in astronomical observational work. Some of these characteristics may even explain currently unidentified features.

In IR spectra of AGB stars, the 11-μm feature is attributed to SiC; it is most commonly an emission feature, but can also be an absorption feature when spectra are collected from stars with optically thick circumstellar envelopes. Laboratory IR spectroscopic work on synthetic SiC by Speck et al. (1999) and Clement et al. (2003) found that, contrary to previous studies that used a KrB correction factor (Borghesi et al. 1983, 1985; Mutschke et al. 1999), the 11-μm feature in spectra from C-stars best matches β-SiC (3C) rather than α-SiC (non-3C). This is consistent with the original TEM studies of presolar SiC (Daulton et al. 2003) in which the 3C polytype was found to be, by far, the most common. However, grain size and shape, degree of agglomeration of the grains, and the impurity content (namely N) seem to be more influential than polytype on the IR spectra (Andersen et al. 1999; Mutschke et al. 1999; Clement et al. 2003). This could mean that large, polycrystalline, heavily-defected, and/or void-rich grains might have IR spectra with a greater shift in the SiC feature (normally at ~11 μm) than grains with unusual polytypes. The presence of minor element heterogeneities and subgrains, both forms of impurities at coarse enough resolution, within SiC dust may also have an effect on the IR spectra. In fact, features within IR spectra of the same stellar object have been observed to vary over time, such as the 10-μm silicate feature (e.g., Monnier et al. 1998; Niyogi et al. 2011). Such observations are consistent with what might be expected with the formation of subgrains from condensation and exsolution as well as changes in the minor element contents over time as a grain crystallizes. Although subgrains may not be large or abundant enough for their IR features to be visible in observational data currently available, they may affect the 11-μm feature of SiC. Additionally, future improvements to the sensitivity of IR spectroscopy for astronomical observations may one day allow for the identification of these subgrains. Given that the mineralogy of the subgrains observed in this study varies, it is worthwhile to discuss IR features for each of the phases. Subgrains observed in this study include ZrC, TiC, TiN, AlN, Fe-bearing phases (metal, carbides, and/or silicides), and CaS.

Laboratory studies on synthetic ZrC observed broad 9.4-μm and 12.4-μm absorption features (Kimura & Kaito 2003). Astronomical observations have not yet identified ZrC (A. Speck, personal communication, 2020 May 21); however, this is not surprising, given that they are expected to be in low abundance as condensates. Laboratory studies of TiC nanocrystals noted a 20.1-μm feature (von Helden et al. 2000). The fact that IR spectra of bulk TiC do not show this feature implies that, if the feature is related to TiC, grain size plays a major role in its presence (Henning & Mutschke 2001). It is also useful to note that the 20.1-μm feature has been matched to SiC dust from astronomical observations (Speck & Hofmeister 2004), so perhaps SiC grains with small TiC subgrains are, in fact, responsible for the feature.

Laboratory spectra of TiN do not show strong features; this lack of features, as well as the low abundance predicted for TiN in circumstellar envelopes of C stars, could explain the fact that TiN has not been identified by astronomical observations (Pitman et al. 2006). A 10-μm and 14.1-μm feature have both been attributed to AlN; the former from modeling of spectra from luminous blue variable star η Carinae (Morris et al. 2017), and the latter from laboratory work on synthetic nitrides (Pitman et al. 2006). TEM analyses from previous studies and the current study have shown that Al and N are often present in SiC as impurities, so that they can occur as Al,N-rich SiC in addition to discrete AlN subgrains. Laboratory IR spectroscopic work on SiC with variable Al and N contents would be most useful in determining what features are likely for Al,N-rich SiC dust.

For the Fe-bearing subgrains in SiC grains, Fe metal, carbides, and silicides are of interest. Modeling of IR spectra from the luminous blue variable star η Carinae inferred the presence of Fe metal around the star, although Fe metal has a flat spectrum and, as such is not uniquely constrained; additionally, the fit required the presence of silicates (Morris et al. 2017). In addition to O abundances greater than expected for the formation of SiC (C/O ratio <1 as the presence of silicates indicates), luminous blue variable stars are very high-mass stars and are unlikely to represent the stellar sources of the MS, Y, or X grains. Laboratory studies of synthetic Fe carbide ($Fe_3C$) by Nuth et al. (1985) found that $Fe_3C$ is largely featureless in the range of 15.63–125 μm. Synthetic $FeSi_2$ spectra, on the other hand, show several features in the 20 μm to 40 μm range, including features at ~23, 26, 28, 32, and 38 μm (Lefki et al. 1991; Fenske et al. 1996). Lastly, IR spectral studies from laboratory analyses of CaS identified absorption features near 40 μm and also possible contributions to the 30 μm feature, most often attributed to MgS (Nuth et al. 1985).

In short, structural features, minor element heterogeneities, and subgrains could potentially influence IR spectra with implications for astronomical observations of circumstellar environments. Future work determining the effect of each of these factors would be useful for better matching observations from TEM studies of presolar SiC grains to the dust from their stellar sources.

## 5. CONCLUSIONS

The crystal structures, elemental compositions, and subgrain contents of presolar grains retain a record of their condensation histories in circumstellar environments. Prior electron diffraction and high-resolution TEM studies have shown that presolar SiC is predominantly cubic 3C (β-SiC), which reflects condensation at the lowest energy conditions. However, our detailed TEM analyses of eight select presolar SiC grains (4 MS, 1 Y, and 3 X), reveal structural and elemental variations that indicate more complex histories than homogenous gas-to-dust condensation in monotonically cooling circumstellar envelopes. The grains display a diversity of structures (single crystal domain 3C to complex intergrown grains with higher order hexagonal polytypes), minor element contents (Al,(Mg,)N rich to poor), and subgrain phases (AlN, TiC, TiN, ZrC, CaS, and/or Fe metal/carbides/silicides). Voids appear to be common. The C/O ratios of grains' progenitor circumstellar environments are inferred to range from 0.98 to ≥1.03 based on the textures and compositions of the grains. Additionally, the selected grains in this study provide evidence for temperature cycling and/or changes in pressure.

Given this diversity, understanding the effects of porosity, Al and N substitution in SiC, and the presence of subgrains of varying compositions on the 11-μm feature, and IR spectra more generally, is essential for better interpreting the astronomical observational data of C-rich AGB stars. We acknowledge that the eight grains featured in this study represent a small data set and, as such, cannot provide information on statistically significant trends or correlations. Additional coordinated chemical and structural studies of presolar SiC grains are required, especially given the advanced analytical techniques now at our disposal.

## Acknowledgements

S.A.S thanks Angela Speck for helpful discussion of IR spectroscopic studies of SiC from an astronomical perspective. This research was supported at NRL by NASA Emerging Worlds grants 80HQTR19T0038 and NNH16AC42I (RMS), 80NSSC20K0387 (NL), and NNX17AE28G (LRN).

**Table 1.** Isotopic compositions, μ-Raman, and TEM data from presolar SiC grains

| Grain | Group | $^{12}C/^{13}C$ | $^{14}N/^{15}N$ | $\delta^{29}Si/^{28}Si$ (‰) | $\delta^{30}Si/^{28}Si$ (‰) | TO center ($cm^{-1}$) | TO width ($cm^{-1}$) | Raman polytype | TEM polytype | Xl domains | Size (nm)[b] | Stacking faults | Voids | Subgrains |
|---|---|---|---|---|---|---|---|---|---|---|---|---|---|---|
| G312 | MS | 71 ± 2 | 380 ± 10 | 40 ± 10 | 30 ± 10 | 799.1 ± 0.2 | 9 ± 1 | 3C | 3C | 3 | 740 | Yes | Yes | TiC, AlN |
| G648 | MS | 51 ± 1 | 800 ± 100 | 70 ± 10 | 60 ± 10 | 791.9 ± 0.4 | 17.8 ± 0.9 | 3C or non-3C | 3C | 1? | 1160 | No | NA | NA |
| G620 | MS | 64 ± 2 | 1500 ± 90 | 10 ± 20 | 30 ± 20 | 800.9 ± 0.5 | 27 ± 6 | 3C | 3C | 2 | 710 | No | Yes | TiN |
| G619[a] | MS | 81 ± 2 | 490 ± 30 | -50 ± 30 | 10 ± 50 | 776.3 ± 0.6<br>763.2 ± 0.2 | 8.9 ± 0.7<br>6.5 ± 0.2 | Intergrowth or non-3C | 2H-4H | 1 | 670 | No | Yes | TiC, Fe-m, CaS |
| G670 | Y | 112 ± 3 | 630 ± 70 | -30 ± 10 | -1 ± 20 | 799.8 ± 0.3 | 15.9 ± 0.5 | 3C | 3C | 2 | 640 | Yes | Yes | TiC |
| G506 | X | 62.7 ± 0.4 | 100 ± 4 | -40 ± 20 | -140 ± 10 | 794 ± 1 | 24 ± 2 | 3C or non-3C | 3C-8H | >7 | 1230 | Yes | Yes | TiC, ZrC |
| G674 | X | 120 ± 3 | 87.4 ± 0.9 | -187 ± 8 | -290 ± 8 | 792.6 ± 0.2 | 10.8 ± 0.2 | 3C or non-3C | 3C-2H-10H-14H | >13 | 1000 | Yes | Yes | Fe-m, Ni-m, FeSi, $Fe_3C$ |
| G1036 | X | 202 ± 7 | NA | -410 ± 10 | -570 ± 20 | 763 ± 1 | 82 ± 7 | non-3C | 3C-2H | 4 | 650 | Yes | Yes | None |

Errors are 1σ; MS – mainstream, NA – not available, TO center – Raman TO peak center, TO width – Raman TO peak width measured as FWHM, Xl domains – number of crystal domains, Fe-m – iron metal, Ni-m – nickel metal

[a]Grain has multiple peak centers in its Raman spectrum

[b]Calculated as the geometrical mean

**Table 2.** TEM EDX minor element (atomic % and ratio) compositions of the SiC in the presolar SiC grains

| Grain | Group | Description | N | Mg | Al | Mg/Al | N/Si | Al/Si |
|---|---|---|---|---|---|---|---|---|
| G312 | MS | SiC | 0.28 ± 0.02 | bdl | 0.15 ± 0.04 | … | 0.005 ± 0.001 | 0.003 ± 0.001 |
| | | Stacking faults | 0.76 ± 0.06 | bdl | 0.26 ± 0.05 | … | 0.013 ± 0.001 | 0.004 ± 0.001 |
| G620 | MS | SiC | 0.55 ± 0.07 | bdl | bdl | … | 0.012 ± 0.002 | … |
| | | Al,N rich | 4.6 ± 0.2 | bdl | 2.3 ± 0.2 | … | 0.117 ± 0.006 | 0.058 ± 0.005 |
| G619 | MS | SiC | 0.52 ± 0.08 | bdl | 0.11 ± 0.05 | … | 0.012 ± 0.002 | 0.003 ± 0.001 |
| | | Al,N rich | 3.0 ± 0.2 | bdl | 2.1 ± 0.2 | … | 0.071 ± 0.004 | 0.050 ± 0.004 |
| G670 | Y | SiC | 0.48 ± 0.07 | bdl | bdl | … | 0.009 ± 0.001 | … |
| G506 | X | SiC | bdl | bdl | 0.16 ± 0.05 | … | … | 0.002 ± 0.001 |
| G674 | X | SiC | 1.9 ± 0.1 | 0.63 ± 0.08 | 1.0 ± 0.1 | 0.6 ± 0.1 | 0.035 ± 0.002 | 0.019 ± 0.002 |
| | | Al,Mg,N rich | 3.0 ± 0.2 | 0.7 ± 0.1 | 1.4 ± 0.2 | 0.5 ± 0.1 | 0.059 ± 0.004 | 0.027 ± 0.003 |
| G1036 | X | SiC | 5.0 ± 0.2 | 1.7 ± 0.1 | 2.2 ± 0.2 | 0.77 ± 0.08 | 0.100 ± 0.004 | 0.043 ± 0.003 |

Errors are 1σ and calculated using uncertainty propagation for the ratios; MS – mainstream, Stacking faults – regions of high-density stacking faults in the grain, Al,(Mg,)N rich – averages of minor element-rich regions in EDX maps, bdl – below detection limit, ... Indicates that at least one element in the ratio was below detection limit

**Table 3.** Textural and compositional (atomic ratio) information on select subgrains in presolar SiC grains

| Grain | Group | Subgrain[a] | Phase[b] | Shape[c] | Size (nm)[d] | Void assoc.[e] | V/Ti | Fe/Ti | Ni/Ti | Zr/Ti | Mo/Ti | Fe/Ni | Zr/Mo |
|---|---|---|---|---|---|---|---|---|---|---|---|---|---|
| G312 | MS | L1 | AlN | Equant a | 20 | Yes | … | … | … | … | … | … | … |
| | | M1 | TiC | Elongate | 30 | No | 0.11 ± 0.03 | … | … | 0.07 ± 0.08 | … | … | … |
| | | M2 | TiC | Equant a | 10 | No | 0.09 ± 0.05 | … | … | … | … | … | … |
| | | M3 | TiC | Elongate | 20 | No | … | … | … | 0.1 ± 0.2 | … | … | … |
| | | M4 | AlN | Elongate | 20 | No | … | … | … | … | | | |
| | | R1 | TiC | Elongate | 60 | No | 0.10 ± 0.03 | … | … | 0.06 ± 0.06 | … | … | … |
| | | R2 | TiC | Elongate | 30 | No | 0.15 ± 0.06 | … | … | 0.1 ± 0.2 | … | … | … |
| | | R3 | TiC | Elongate | 20 | No | … | … | … | … | … | … | … |
| G620 | MS | 1 | TiN | Equant s | 40 | No | 0.025 ± 0.003 | … | … | 0.028 ± 0.002 | … | … | … |
| G619 | MS | 1 | TiC | Elongate | 30 | No | 0.11 ± 0.03 | 0.05 ± 0.02 | … | … | … | … | … |
| | | 2 | TiC | Elongate | 20 | No | 0.03 ± 0.01 | 0.08 ± 0.02 | 0.08 ± 0.02 | … | … | 1.0 ± 0.3 | … |
| | | 3 | Fe-m | Equant a | 10 | Yes | … | 90 ± 40 | 5 ± 2 | 1 ± 1 | 1 ± 1 | 20 ± 4 | 1 ± 2 |
| | | 4 | TiC | Elongate | 20 | No | … | 0.03 ± 0.01 | … | 0.09 ± 0.05 | … | … | … |
| | | 5 | TiC | Elongate | 20 | Yes | … | … | … | … | … | … | … |
| | | 6 | TiC | Equant s | 20 | Yes | 0.14 ± 0.02 | 0.06 ± 0.01 | 0.018 ± 0.006 | 0.08 ± 0.03 | 0.02 ± 0.01 | 3 ± 1 | 4 ± 3 |
| | | 7 | TiC | Equant a | 10 | No | 0.05 ± 0.01 | 0.05 ± 0.01 | 0.03 ± 0.01 | 0.06 ± 0.03 | 0.02 ± 0.02 | 1.9 ± 0.9 | 3 ± 3 |
| G670 | Y | 1 | TiC | Elongate | 30 | Yes | … | … | … | … | … | … | … |
| | | 2 | TiC | Elongate | 20 | Yes | … | … | … | … | … | … | … |
| | | 3 | TiC | Elongate | 20 | No | … | … | … | … | … | … | … |
| | | 4 | TiC | Elongate | 50 | Yes | 0.13 ± 0.04 | … | … | … | … | … | … |
| G506 | X | 1 | TiC | Elongate | 20 | Yes | 0.18 ± 0.02 | … | … | 0.08 ± 0.02 | … | … | … |
| | | 2 | ZrC | Equant a | <10 | No | … | … | … | 1.8 ± 0.4 | … | … | … |
| | | 3 | ZrC | Equant a | <10 | No | … | … | … | 2.3 ± 0.5 | 0.2 ± 0.1 | … | 10 ± 10 |
| | | 5 | TiC | Equant s | <10 | No | … | … | … | 0.5 ± 0.1 | 0.05 ± 0.05 | … | 9 ± 9 |
| | | 6 | ZrC | Equant e | <10 | No | … | … | … | 3.5 ± 0.9 | … | … | … |
| | | 7 | ZrC | Equant a | <10 | Yes | … | … | … | 5 ± 1 | … | … | … |
| | | 8 | ZrC | Equant a | <10 | Yes | … | … | … | 3.0 ± 0.7 | … | … | … |
| | | 9 | ZrC | Equant a | <10 | No | … | … | … | 2.0 ± 0.5 | … | … | … |
| | | 10 | TiC | Equant s | 10 | No | … | … | … | 0.22 ± 0.07 | 0.04 ± 0.04 | … | 5 ± 5 |
| | | 11 | ZrC | Equant a | <10 | No | … | … | … | 2.1 ± 0.5 | … | … | … |
| | | 12 | ZrC | Equant a | <10 | No | … | … | … | 1.7 ± 0.5 | … | … | … |

| Grain | Group | Subgrain | Phase[a] | Shape | Size (nm)[b] | Void assoc.[c] | V/Ti | Fe/Ti | Ni/Ti | Zr/Ti | Mo/Ti | Fe/Ni | Zr/Mo |
|---|---|---|---|---|---|---|---|---|---|---|---|---|---|
| | | 13 | ZrC | Equant a | <10 | No | … | … | … | 1.6 ± 0.4 | 0.2 ± 0.2 | … | 7 ± 5 |
| | | 14 | ZrC | Equant a | <10 | No | … | … | … | 5 ± 3 | … | … | … |
| | | 15 | ZrC | Equant a | <10 | No | … | … | … | 1.5 ± 0.4 | 0.2 ± 0.1 | … | 10 ± 8 |
| | | 16 | ZrC | Equant a | <10 | No | … | … | … | 2 ± 1 | … | … | … |
| G674 | X | 1 | Fe-m | Equant a | 20 | Yes | … | 60 ± 30 | … | … | … | … | … |
| | | 2 | Fe-m | Equant s | 20 | Yes | … | 40 ± 10 | … | … | … | … | … |
| | | 3 | Fe-m | Equant s | 30 | Yes | … | 90 ± 40 | … | … | … | … | … |
| G674 | X | 4 | Fe-m | Equant a | 20 | Yes | … | 1.0 ± 0.1 | … | … | … | … | … |
| | | 5 | Fe-m | Elongate | 10 | Yes | … | 20 ± 10 | … | … | … | … | … |
| | | 6 | FeSi | Equant a | 10 | No | … | 10 ± 4 | … | … | … | … | … |
| | | 7 | Ni-m | Elongate | 10 | Yes | … | 6 ± 3 | 20 ± 10 | … | … | 0.29 ± 0.06 | … |
| | | 8 | FeSi | Equant a | 10 | No | … | 12 ± 6 | … | … | … | … | … |
| | | 9 | Fe-m | Equant a | 10 | Yes | … | … | … | … | … | … | … |
| | | 10 | Fe-m | Equant s | 20 | No | 1.2 ± 0.6 | 15 ± 5 | … | … | … | … | … |
| | | 11 | $Fe_3C$ | Equant a | <10 | No | … | 20 ± 10 | 4 ± 3 | … | … | 5 ± 1 | … |

[a]Figures A9–A10 show images of the subgrains for which data was collected
[b]As determined with compositional information
[c]Equant refers to equidimensional grains (a – anhedral, s – subhedral, and e – euhedral)
[d]Calculated as the geometrical mean
[e]Subgrains in association with voids
Errors are 1σ and calculated using uncertainty propagation for the ratios; MS – mainstream, Fe-m – iron metal, Ni-m – nickel metal ... Indicates that at least one element in the ratio was below detection limit

# 6. APPENDIX

## 6.1 Appendix Figures

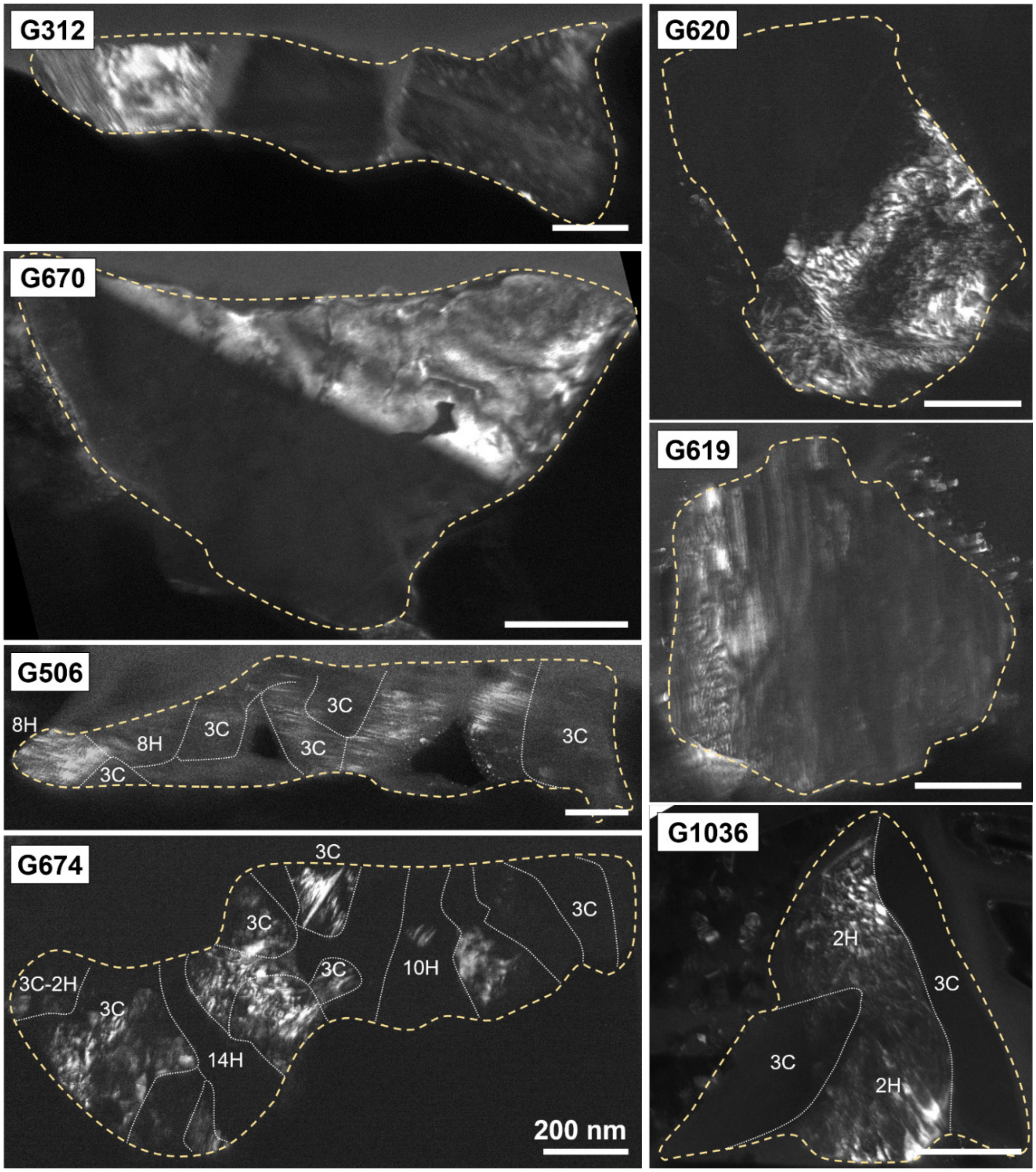

**Figure A1.** Representative "dirty" dark field (DDF) images of MS grains G312, G620, G619; Y grain G670; and X grains G506, G674, and G1036. All scale bars are 200 nm. G648 was lost during sample exchange before any DDF images were collected. Together with SAED patterns, several DDF images were used to determine the locations and numbers of crystal domains in each grain. Grains with well-defined domains are labeled with their polytype for each.

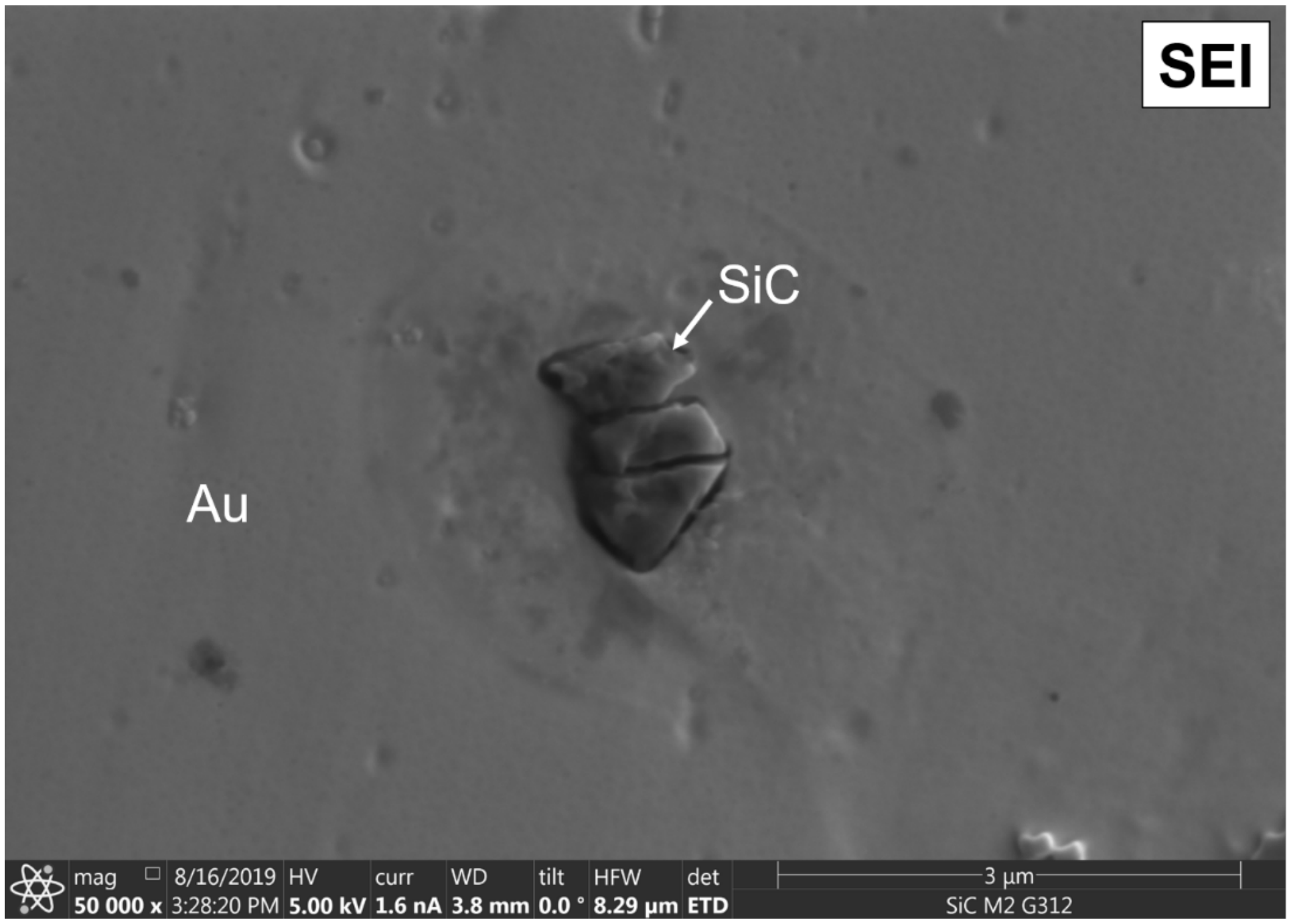

**Figure A2.** Secondary electron image (SEI) of MS grain G312 embedded in Au foil. The three fragments of the grain, separated by cracks, are clearly visible. The appearance of the grain implies that it was fractured when it was pressed into the Au foil. The FIB section was cut roughly in a vertical orientation, so that the topmost fragment here corresponds to the leftmost in the TEM images.

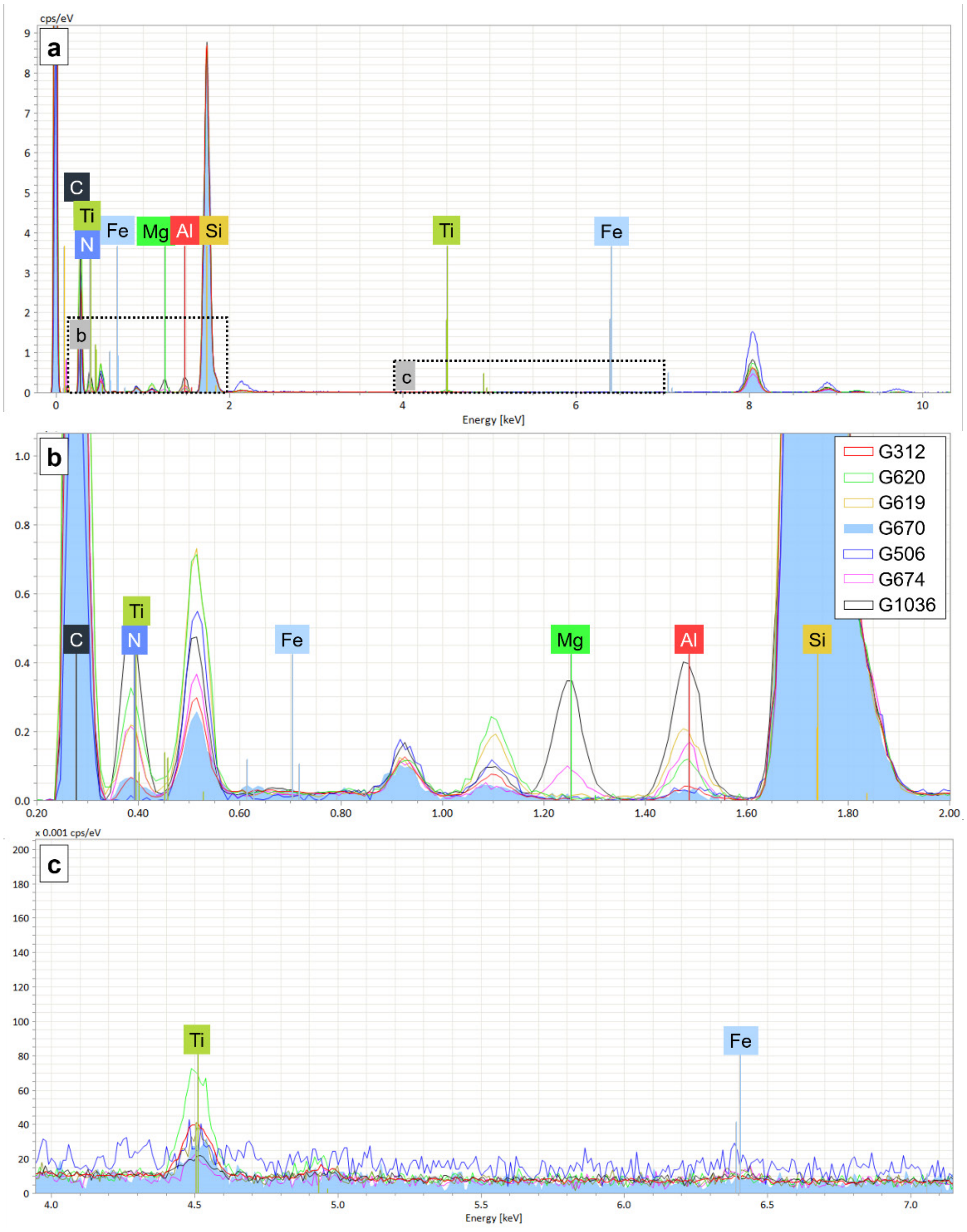

**Figure A3.** EDX spectra of overall presolar SiC MS grains G312, G620, G619; Y grain G670; and X grains G506, G674, and G1036, all scaled to the Si peak. The spectra from the zero peak to ~10 keV are shown in (a). The differences in the minor elements better distinguished in (b) for Al, Mg, and N and (c) for Ti and Fe. Unlabeled peaks represent contamination either from sample preparation or components within the microscope. Al,(Mg),N heterogeneities in G620, G619, G674, and G1036 are responsible for the high minor element contents in the spectra of those grains. The discrete AlN subgrains in G312 do not appear to contribute to high minor element contents in the overall grain. Al, Mg, and N abundances are low in G506 and G670, consistent with their lack of Al,(Mg),N impurities.

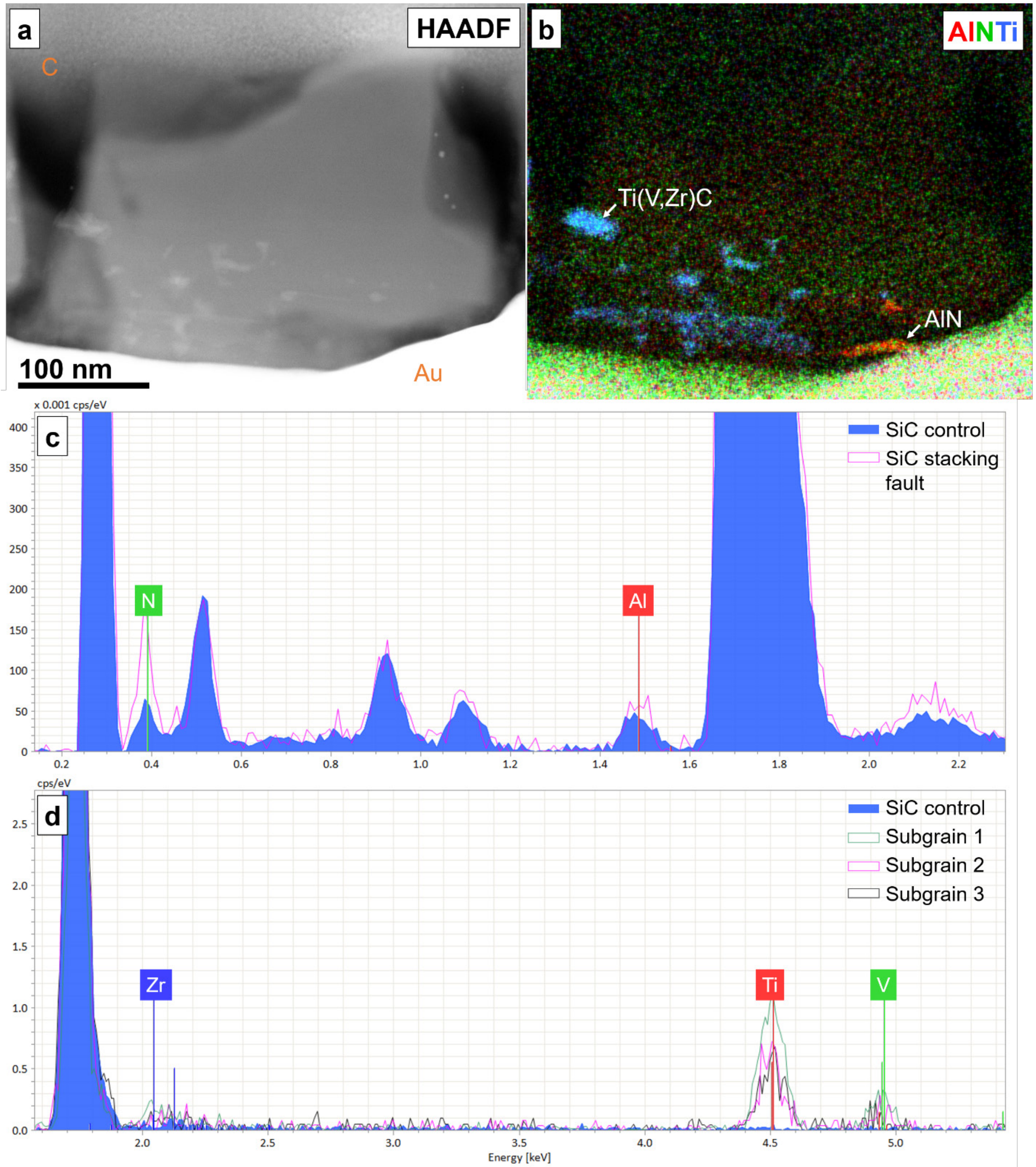

**Figure A4.** STEM HAADF image, EDX map (AlNTi), and EDX spectra of MS grain G312. (a) The Au foil, supporting the SiC, and the protective C strap are labeled in the STEM HAADF image. (b) The composite EDX map shows the presence of TiC subgrains (blue) and discrete AlN subgrains (red-green). EDX spectra are scaled to the zero peak (not visible). (c) Example spectra of SiC from regions of low disorder (SiC control) and high disorder (SiC stacking fault) show a difference between the N and Al contents. (d) Example spectra of subgrains (Subgrains 1–3) and adjacent, representative SiC (SiC control) illustrate the differences in Zr, Ti, and V abundances.

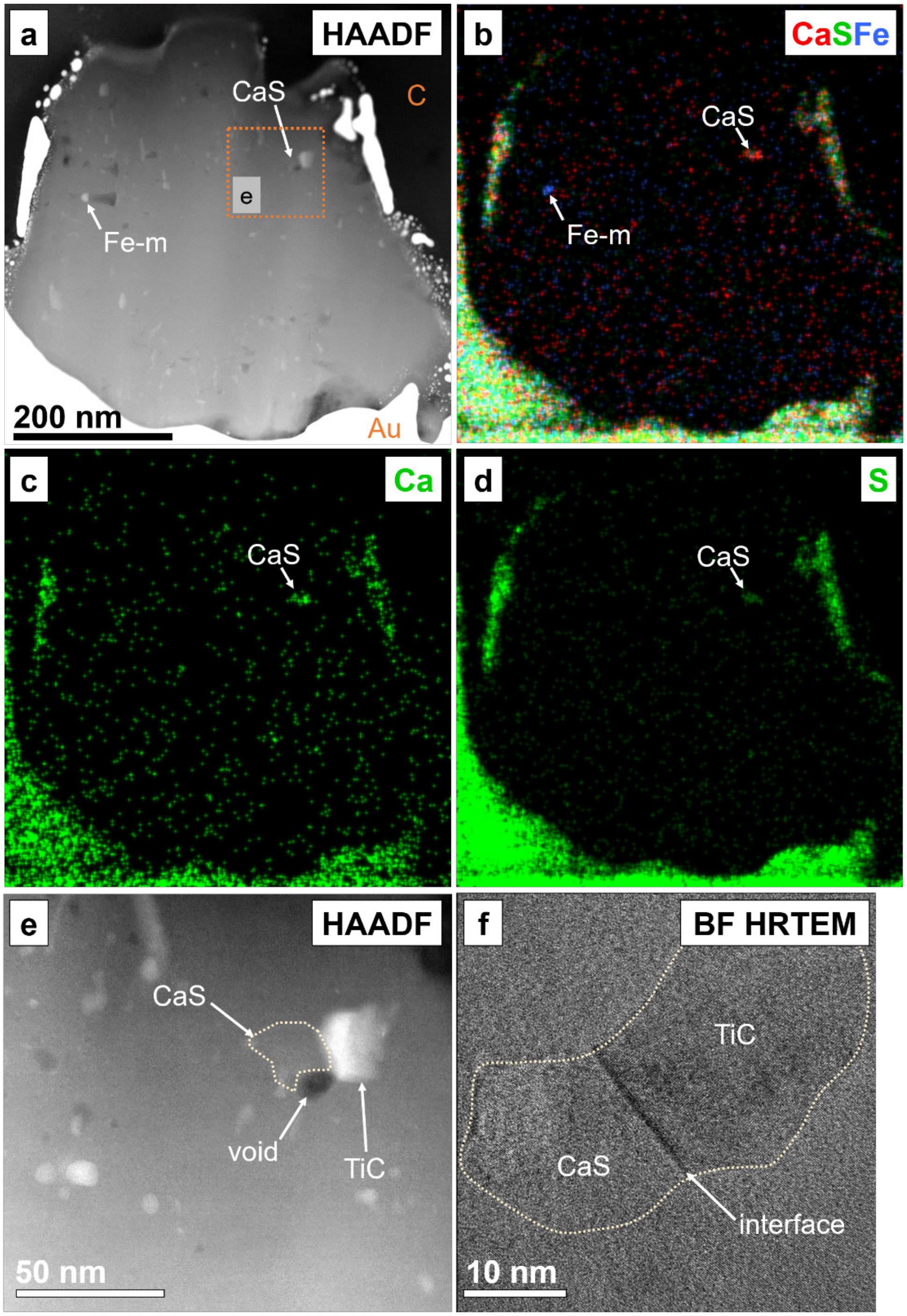

**Figure A5.** STEM HAADF images, BF HRTEM images, and EDX maps of MS grain G619. (a) The Au foil, supporting the SiC, and the protective C strap are labeled. (b) The composite EDX map (CaSFe) shows the presence of the Fe-metal subgrain (blue) and a CaS subgrain (red). The Ca (c) and S (d) EDX maps show the localized enrichment of each element more clearly. The low signal-to-noise ratio for Ca results in a map with more noise. (e) The higher magnificent STEM HAADF image shows the CaS grain adjacent to a TiC grain and void. (f) The BF HRTEM image illustrates the sharp interface between the CaS and TiC subgrains.

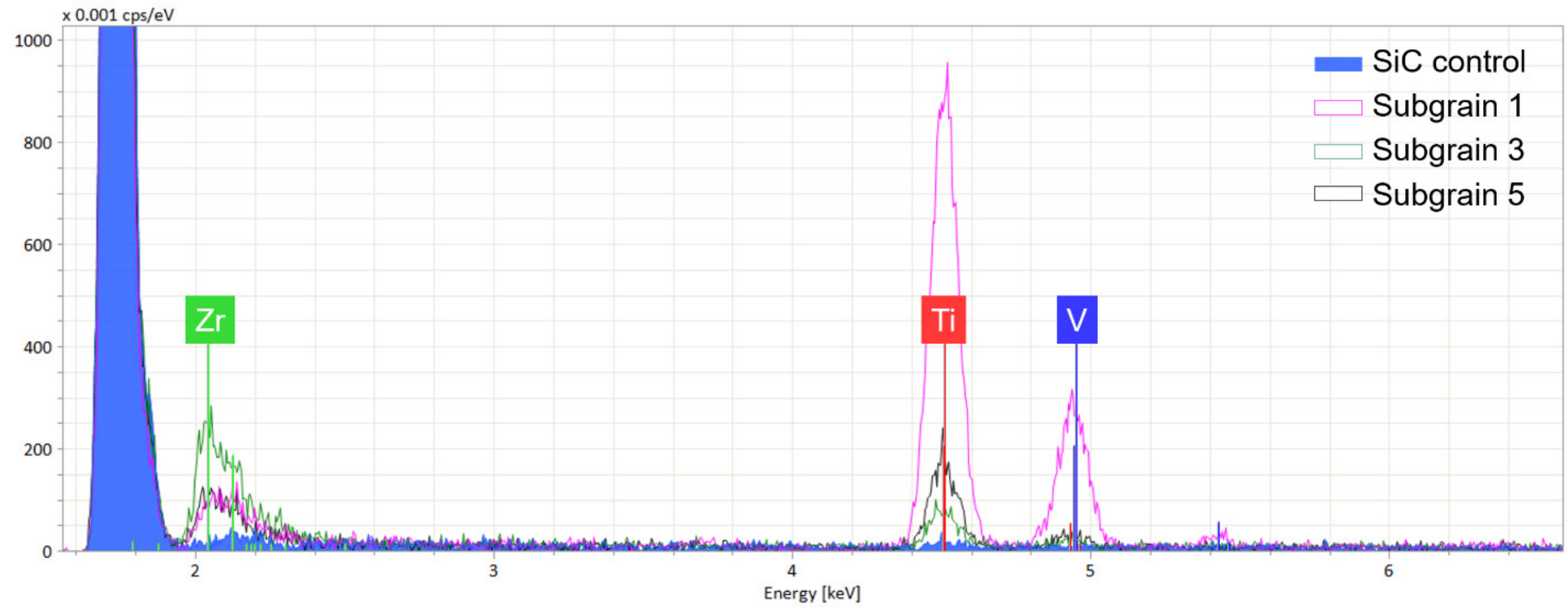

**Figure A6.** EDX spectra of X grain G506, scaled to the zero peak (not visible). Example spectra of subgrains (Subgrains 1, 3, and 5) and adjacent, representative SiC (SiC control) illustrate the differences in Zr, Ti, and V abundances. In quantifying the spectra of the subgrains, the Zr abundances of the SiC control were subtracted from the Zr abundances of the subgrains.

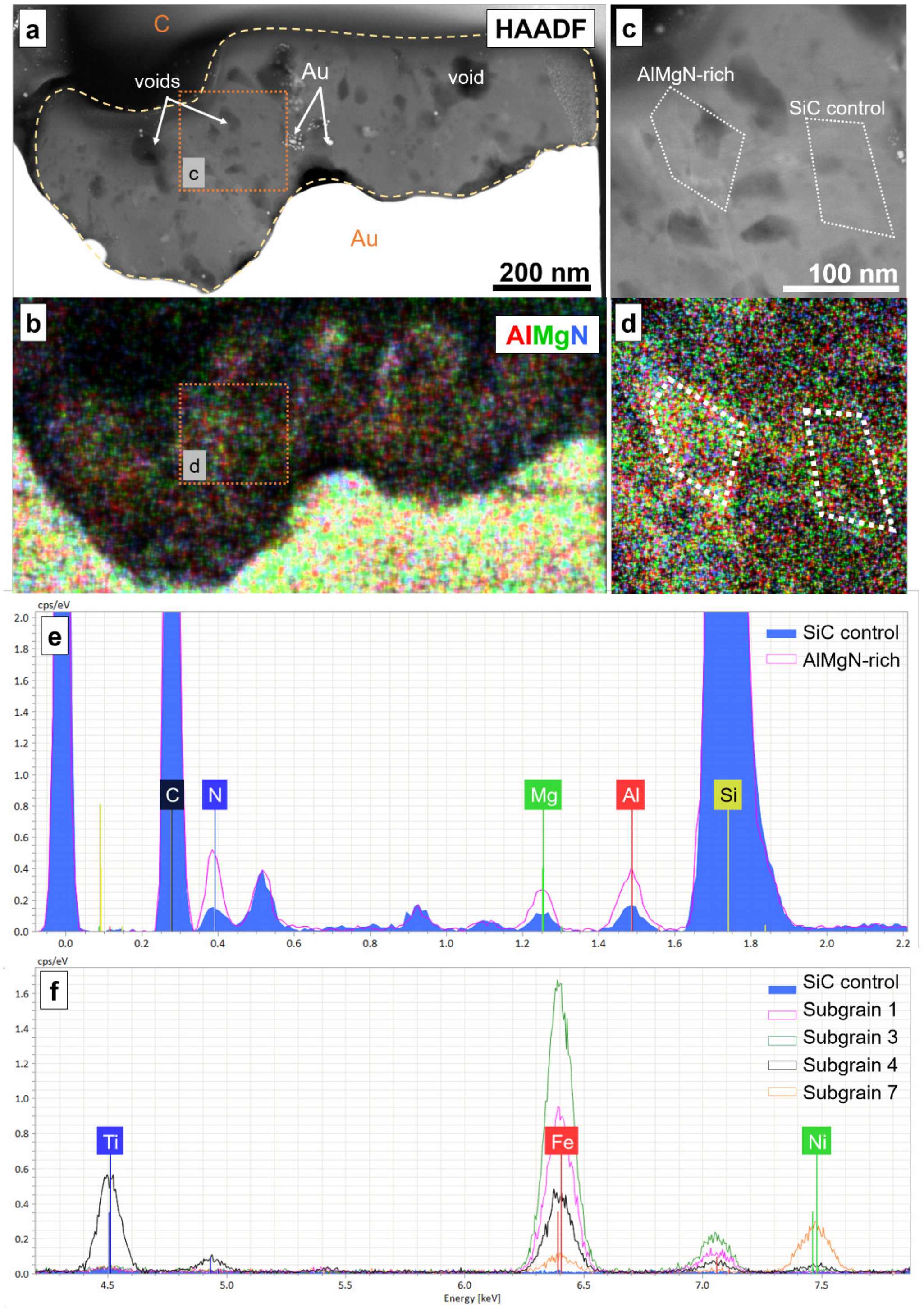

**Figure A7.** STEM HAADF images, EDX maps (AlMgN), and EDX spectra of X grain G674. (a) The Au foil, supporting the SiC, and the protective C strap are labeled. The variation in the Al, Mg, and N contents of the grain are shown in the high magnification EDX map (d) and the spectra (e), scaled to the zero peak, from the regions of interest indicated in (c). (f) Example spectra of subgrains (Subgrains 1, 3,

4, and 7) and adjacent, representative SiC (SiC control) illustrate the differences in Ti, Fe, and Ni abundances.

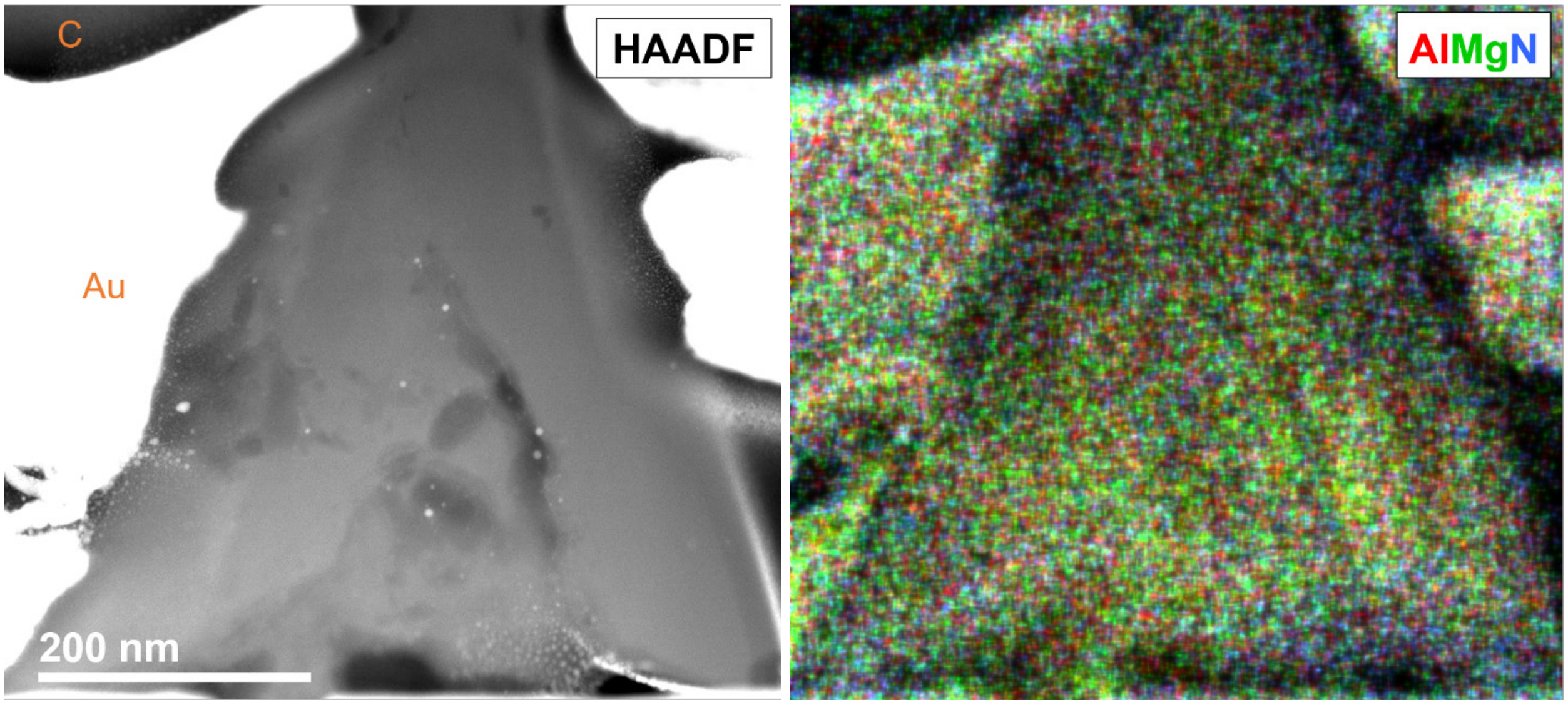

**Figure A8.** STEM HAADF image and EDX map (AlMgN) of X grain G1036. The Au foil, supporting the SiC, and the protective C strap are labeled in the STEM HAADF image. The bright white material in the HAADF image is Au contamination from FIB sample preparation. The composite EDX map shows some variation in the Al, Mg, and N contents of the grain, with the bottom portions of the grain having higher minor element contents.

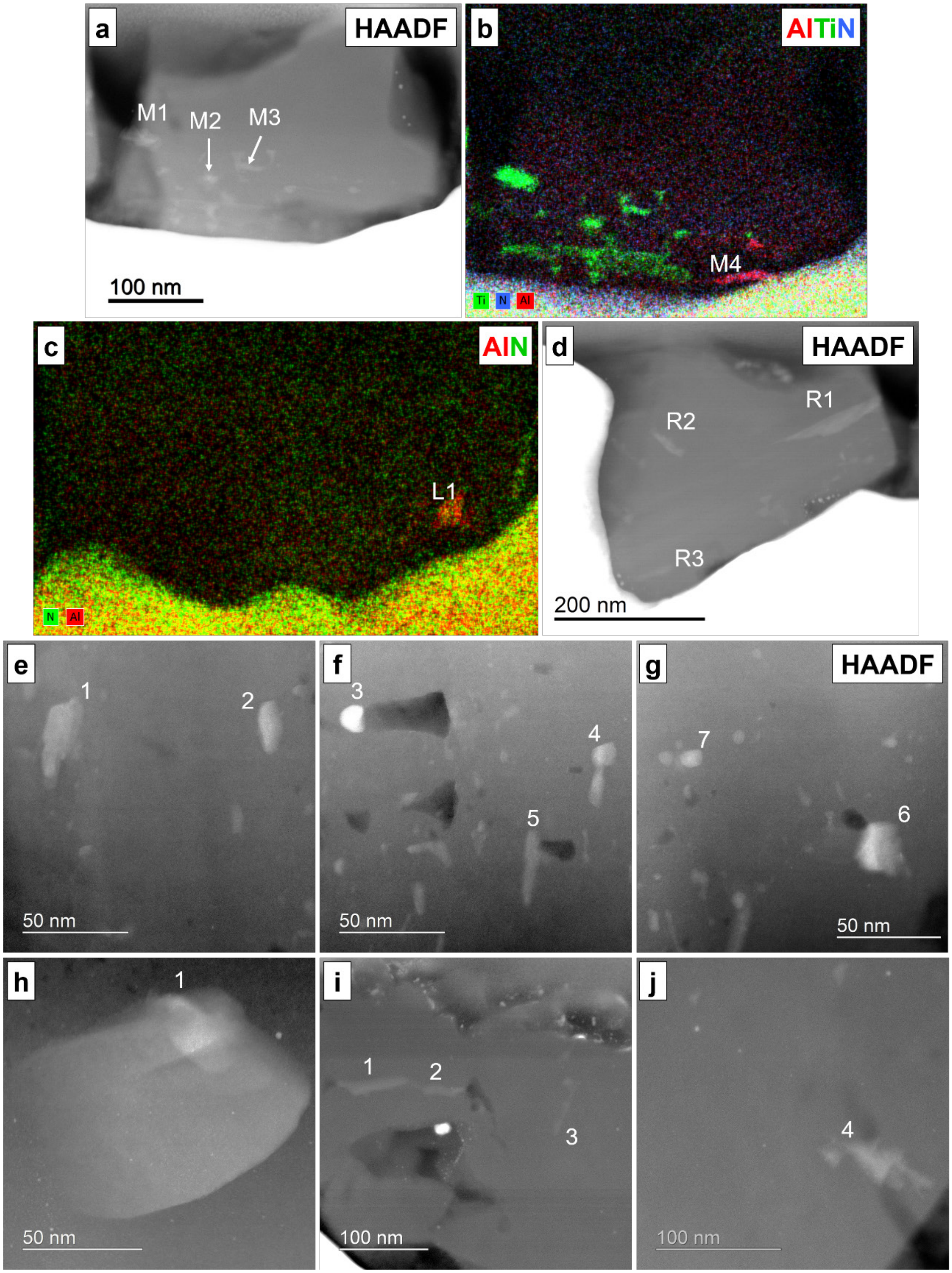

**Figure A9.** STEM HAADF images and EDX maps of MS and Y grains showing the subgrains for which compositional data is presented in Table 3: (a–d) MS G312, (e–g) MS G619, (h) MS G620, and (i–j) Y G670. AlTiN (b) and AlN (c) composite EDX maps are necessary to show the presence of AlN subgrains.

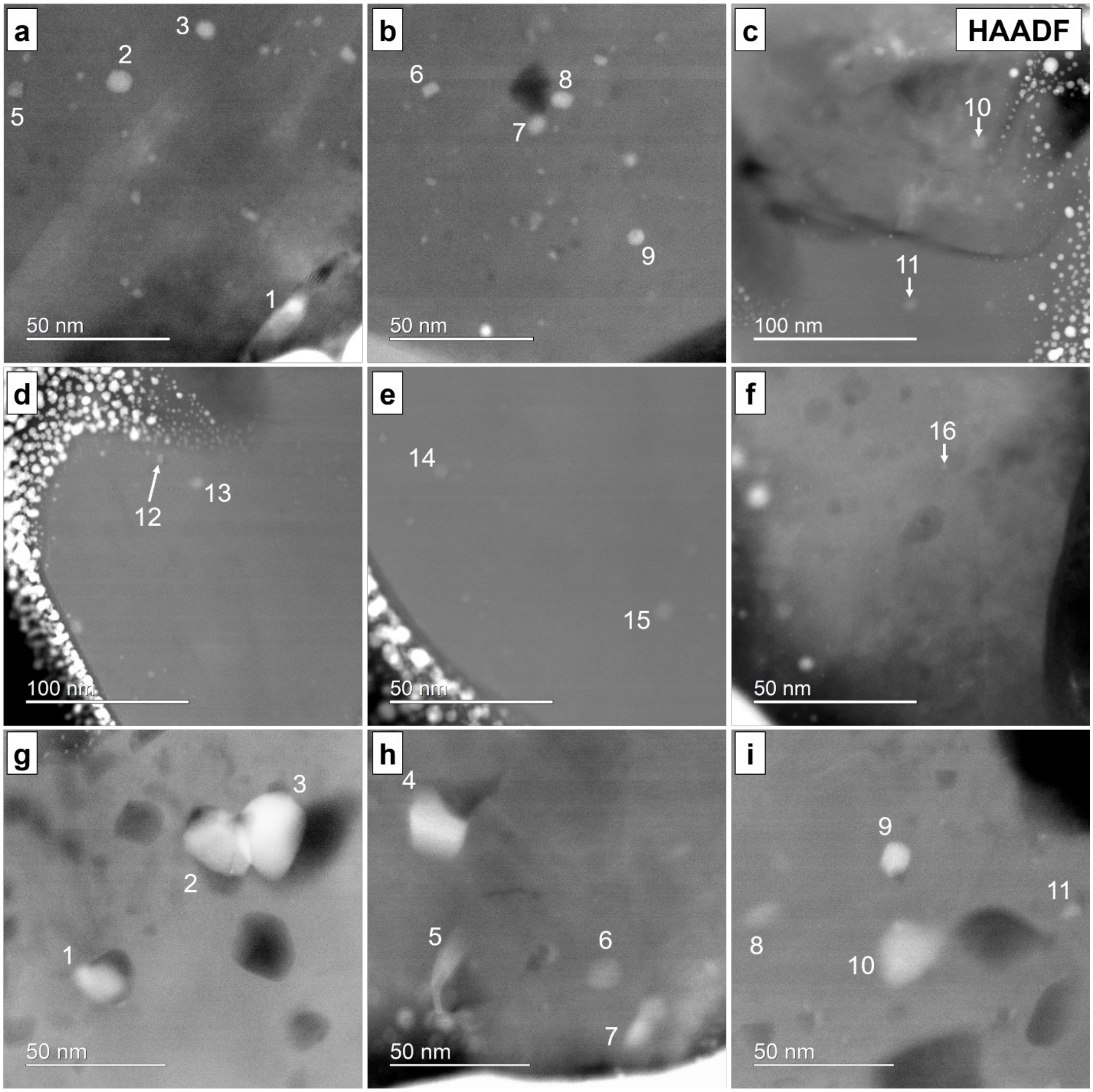

**Figure A10.** STEM HAADF images of X grains showing the subgrains for which compositional data is presented in Table 3: (a–f) X G506 and (g–i) X G674. Bright white, spherical material in G506 is Au contamination from FIB sample preparation.

## 6.2 Appendix Tables

Table A1: Grain, Void, and Subgrain Sizes
Diameter measurements and geometric means of grains, voids, and subgrains

Table A2: EDX Data
EDX data obtained from quantification of spectra for SiC and subgrains. The SiC control spectra used to determine relative enrichments and depletions of elements between the subgrains and adjacent SiC are also included

Table A3: Crystallographic Data
Literature values for phases analyzed for SAED pattern indexing, including crystal system, space group, unit cell parameters, and sources of the data

Table A4: SAED Indexing Summary
Measured and theoretical d-spacings, angles, and goniometer tilts for all domains/grains

Tables A5–A7: SAED Indexing of Higher-Order Polytypes in G619, G506, and G674
More detailed calculations of the crystal domains with higher order (non-3C,2H) polytypes